\documentclass[a4paper, 11pt]{article}

\usepackage[english]{babel}
\usepackage[utf8x]{inputenc}
\usepackage[T1]{fontenc}

\usepackage[a4paper,top=3cm,bottom=2cm,left=3cm,right=3cm,marginparwidth=1.75cm]{geometry}
\usepackage{amsfonts,amssymb,amsmath,amscd,latexsym}
\usepackage{tikz}
\usepackage{graphicx}
\usepackage[colorinlistoftodos]{todonotes}
\usepackage[colorlinks=true, allcolors=blue]{hyperref}
\usepackage{natbib}
\usepackage{dsfont}

\usepackage{xspace}

\newtheorem{Remark}{Remark}

\newcommand{\Zcal}{\mathcal{Z}}
\newcommand{\Esp}{\mathbb{E}}
\newcommand{\Espt}{\widetilde{\Esp}}

\newcommand{\Dcal}{\mathcal{D}}
\newcommand{\Mcal}{\mathcal{M}}
\newcommand{\Ncal}{\mathcal{N}}
\newcommand{\Pcal}{\mathcal{P}}
\newcommand{\Var}{\mathbb{V}}

\newcommand{\trans}{{\intercal}}

\newcommand{\ind}{\mathds{1}}
\newcommand{\btheta}{\theta}
\newcommand{\bY}{Y}
\newcommand{\bZ}{Z}
 
\newcommand{\papprox}{\pt_{\bY}}
\newcommand{\papproxs}{\pt_{\bY^{(s)}}}
\newcommand{\pexact}{p_{\bY}}
\renewcommand{\d}{\text{ d}}
\newcommand{\pbar}{\overline{p}}
\newcommand{\gammabar}{\overline{\gamma}}

\newcommand{\bthetahm}{\bZ_h^m,\btheta_h^m}

\newcommand{\bthetahmoins}{\bZ_{h-1},\btheta_{h-1}}
\newcommand{\bthetahmoinsm}{\bZ_{h-1}^m,\btheta_{h-1}^m}
\newcommand{\bthetah}{\bZ_h,\btheta_h}
\newcommand{\bthetak}{\bZ_{k},\btheta_{k}}
\newcommand{\bthetakmoins}{\bZ_{k-1},\btheta_{k-1}}

\newcommand{\bthetaz}{\bZ_{0},\btheta_{0}}
\newcommand{\bthetazh}{\bZ_{0:h},\btheta_{0:h}}
\newcommand{\bthetazm}{\bZ_{0}^m,\btheta_{0}^m}

\newcommand{\bthetazhm}{\bZ_{0:h}^m,\btheta_{0:h}^m}

\newcommand{\bthetazhmoins}{\bZ_{0:h-1},\btheta_{0:h-1}}

\newcommand{\wtilde}{\widetilde{w}}

\newcommand{\wh}{w_h}
\newcommand{\whm}{w_h^m}
\newcommand{\Whm}{W_h^m}
\newcommand{\Wh}{W_h}

\newcommand{\SBMreg}{SBM-reg\xspace}

\newcommand{\alphat}{\widetilde{\alpha}}
\newcommand{\gammat}{\widetilde{\gamma}}
\newcommand{\betat}{\widetilde{\beta}}
\newcommand{\nut}{\widetilde{\nu}}
\newcommand{\thetat}{\widetilde{\theta}}
\newcommand{\ph}{p_h}
\newcommand{\rhoh}{\rho_h}
\newcommand{\qapprox}{\pt_{\bY}}

\newcommand{\alphakl}{{\alpha_{k\ell}}}

\newcommand{\et}{\widetilde{e}}
\newcommand{\nuk}{\nu_k}
\newcommand{\nul}{\nu_\ell}
\newcommand{\nuK}{\nu_K}

\newcommand{\qt}{\widetilde{q}}
\newcommand{\pt}{\widetilde{p}}
\newcommand{\Vt}{\widetilde{V}}
\newcommand{\etaijkl}{\eta_{ij}^{k\ell}}
\newcommand{\tauik}{\tau_{ik}}

\newcommand{\taujl}{\tau_{j\ell}}

\newcommand{\taut}{\widetilde{\tau}}
\newcommand{\tautik}{\taut_{ik}}

\newcommand{\tautjl}{\taut_{j\ell}}

\newcommand{\xij}{x_{ij}}
\newcommand{\Yij}{Y_{ij}}
\newcommand{\Zik}{Z_{ik}}

\newcommand{\Zjl}{Z_{j\ell}}

\newcommand{\gv}{\;|\;}
\newcommand{\PSBMreg}{Poisson SBM-reg\xspace}

\definecolor{darkgreen}{rgb}{0,0.4,0}

\title{Bayesian inference for network Poisson models}
\author{Sophie Donnet, Stéphane Robin, \\ UMR MIA-Paris, AgroParisTech, INRA, Université Paris-Saclay, 75005 Paris, France}

\begin{document}
\maketitle

\begin{abstract}
	This work is motivated by the analysis of ecological interaction networks. Poisson stochastic blockmodels are widely used in this field to decipher the structure that underlies a weighted network, while accounting for covariate effects. 
	Efficient algorithms based on variational approximations exist for frequentist inference, but without statistical guaranties as for the resulting estimates.
	In absence of variational Bayes estimates, we show that a good proxy of the posterior distribution can be straightforwardly derived from the frequentist variational estimation procedure, using a Laplace approximation. We use this proxy to sample from the true posterior distribution via a sequential Monte-Carlo algorithm. As shown in the simulation study, the efficiency of the posterior sampling is greatly improved by the accuracy of the approximate posterior distribution. The proposed procedure can be easily extended to other latent variable models. We use this methodology to assess the influence of available covariates on the organization of two ecological networks, as well as the existence of a residual interaction structure.
\end{abstract}


\section{Introduction \label{sec:intro}}
\subsection{Motivation}

\paragraph{Ecological networks.}
It is now a commonplace that networks provide a natural and convenient framework to depict the interactions between a set of entities. Ecological networks, which aim at describing the interactions between a set of individuals or species constitute an emblematic example  \citep[see][]{PSK16}. Understanding the global organization (or {\it topology}) of such a network then comes at stake, in order to understand the functioning of an ecosystems or to anticipate its response to some environmental change. \\
The present paper is motivated by the analysis of two typical data sets. The first one, first described by \cite{VPD08}, consists of the number of fungal parasites shared across 51 tree species. The aim is too understand both if genetic or geographical similarities between the species contribute to explain the number of parasites they share and to describe the remaining structure that underlies the network and that is not due to the known similarities. The second example \citep[due to][]{RSF15} relates to animal behavior: interactions between individuals are defined by the number of direct contacts between pairs of onagres during a given period of time. Again, covariates (sex, age) have been collected for each individual and the goal is both to assess the effect of these covariates on the intensity of the interactions and to unravel some residual social organisation, that also contributes to structure the network.

\paragraph{(Weighted) Stochastic block-models.}
The stochastic block-model \citep[SBM: ][]{HoL79} has become a popular model in many fields to unravel the latent structure that underlies an observed network. SBM assumes that each entity belongs to a (hidden) group and that the interaction between two entities is ruled by their respective group memberships. The clustering of nodes (e.g. species or individuals) into groups, that can be interpreted as roles in the system, makes SBM attractive for many applications. SBM has been adapted to the specificities of ecological networks. First, the observed interactions are often weighted (or {\it valued}, e.g.: number of contacts between two individuals, number of common parasites). Secondly, covariates describing the (pairs of) individuals are often available. The former specificity led to extend SBM to a relevant emission distribution to handle all the available information, rather than to reduce the information to a simple presence or absence of interaction. The latter specificity suggests to include the effect of covariates in SBM so that to distinguish between their respective effects and the remaining latent structure, which described the unexplained heterogeneity of the network.

\paragraph{Inference of SBMs.}
Because it relies on unobserved variables (the node's membership), SBM is an incomplete data model. But, likewise many statistical model devoted to networks, SBM displays a complex dependency structure that hampers the use of standard inference techniques such as the EM algorithm \citep{DLR77}. Variational approximations \citep{BKM17,WaJ08} are often used to circumvent this complexity, resulting in variational EM (VEM) algorithms. Unfortunately, statistical guaranties about of the resulting estimates are only available for the simplest version of SBM, namely with binary interactions and in absence of covariates,  only in a asymptotic framework. As a consequence, in many situations, the practitioner can resort to an efficient VEM algorithm to get estimates of the parameters of a well-suited weighted SBM model, as well a a pseudo-ICL criterion for model selection. However, the practioner is deprived from any measure of uncertainty of the estimates and from any statistically guaranty about the model selection criterion. Moreover, from a Bayesian perspective,  Variational Bayes approximations \citep{BeG03} are only available when conjugacy properties arise or when some bound for the likelihood can be derived \citep[see e.g.][]{JaJ00,LRO18}. Neither of these cases happen for the Poisson SBM in presence of covariates.

\paragraph{Our contributions.}
In this paper, we consider the Bayesian inference of a weighted SBM where interactions have a Poisson distribution and which include a regression term to account for the effect of covariates. This model was first introduced by \cite{MRV10} and we will refer to it as \SBMreg. We choose this model both because of its interest for many ecological networks and as a proof-of-concept. Our main contribution is to show that one can easily derive an approximation of the posterior distribution from the {\it frequentist} inference carried out by a VEM algorithm. Then, we show how to design a powerful sequential Monte-Carlo (SMC) sampler that takes the proxy of the posterior as an input and returns a (weighted) sample of the true posterior. This results in a grounded Bayesian inference framework that enables us to asses the effects of the covariates or to compare models.

\paragraph{Outline.}
In the rest of the introduction we fix some notations and introduce formally the \SBMreg model. In Section \ref{sec:infer}, we show how to derive an approximate posterior as a by-product of VEM (\ref{sec:VAR}) and we describe the proposed SMC sampler (\ref{sec:SMC}). Section \ref{sec:illust} is first devoted to some simulation studies that demonstrate the efficiency of the proposed approach  (\ref{subsec:simu}) and to the analysis  of  the datasets described above (\ref{subsec:data}).

\subsection{Model}

We consider a dataset describing the interactions between $n$ nodes and denote by $Y_{ij}$ the interaction count between nodes $i$ and $j$ ($1 \leq i, j \leq n$). In the present paper we consider the case where the interaction matrix $\bY = [\Yij]_{1 \leq i, j \leq n}$ is symmetric ($\Yij = Y_{ij}$) and that nodes do not interact with themselves ($Y_{ii} = 0$). We further assume that, for each pair of node $(i, j)$, a $d$-dimensional vector  of covariates $\xij$ is available. Then, like in the classical SBM framework, we assume that $K$ groups of nodes exist and that each node belongs to one and only one group. The \PSBMreg model we consider states that the distribution of the interaction $\Yij$ depends on both the covariate vector  and the groups to which nodes $i$ and $j$ belong.

Formally, we denote $Z_i$ ($1 \leq Z_i \leq K$) the group to which node $i$ belongs  and we assume that the $\{Z_i\}_{1 \leq i \leq n}$ are all independent with distribution 
\begin{equation}\label{eq:Z}
Z_i \sim \Mcal(1; \nu)
\end{equation}
where $\nu = (\nu_1, \dots, \nuK)$ stands for the vector of group proportions (with $\sum_{k=1}^K \nu_k = 1$). Then we assume the interaction weights $\{\Yij\}_{1 \leq i < j \leq n}$ are independent conditionally on the $\{Z_i\}_i$, with distribution
\begin{equation}\label{eq:Y}
\left(Y_{ij} \gv Z_i = k, Z_j = \ell\right) \sim \Pcal(\exp(\alphakl + \xij^\intercal \beta))
\end{equation}
where $\Pcal$ is the Poisson distribution,  $\alphakl$ stands for the interaction term between groups $k$ and $\ell$ and $\beta$ denotes the $d$-dimensional vector of regression coefficients, which encodes the effects of the covariates. The interaction matrix $\alpha = [\alphakl]_{1 \leq k, \ell \leq K}$ is obviously symmetric. Hence the model is parametrized with $\theta = (\nu, \alpha, \beta)$, which consists in $(K-1) + K(K+1)/2 + d$ independent parameters.

\paragraph{Notations}
In what follows, $\bZ$ is the set of latent variables, $p_{\theta}(\bZ)$ is the probability density function (pdf) of the $\bZ$ for a fixed parameter $\btheta$ and $p_{\bZ,\btheta}(Y)$ is the conditional pdf of $\bY$ given $\bZ$ and $\btheta$. Thus, $p_{\theta}(\bY,\bZ) =p_{\btheta}(\bZ) p_{\bZ,\btheta}(Y)  $ is the joint distribution density of $\bY$ and $\bZ$ for a given $\btheta$. We refer to  $\btheta \mapsto  p_{\theta}(\bY,\bZ)$ as the complete likelihood whereas  $ \theta \mapsto p_{\theta}(\bY)   = \sum_{\bZ \in  \mathcal{Z}}  p_{\theta}(\bY,\bZ)  $ is the   likelihood.  The complete  log-likelihood writes
\begin{align*}
 \log p_\theta(Y, Z) = \sum_{i, k} Z_{ik} \log \nu_k + \sum_{i<j} \sum_{k, \ell} Z_{ik} Z_{j\ell} \log  f_{\Pcal}(Y_{ij};e^{\alphakl + \xij^\intercal \beta})
\end{align*}
where $f_{\Pcal}$ is the density of a Poisson distribution.  Moreover, letting $\pi(\btheta)$ denote the prior distribution on $\theta$,
$$
\pexact(\bZ, \btheta) = p(\bZ, \btheta  \mid  \bY) = \frac{\pi(\btheta)p_{\theta}(\bY,\bZ)}{p(\bY)}
$$
is the posterior  distribution of $(\bZ,\btheta)$ where  $p(\bY) = \int_{\theta\in \Theta}  p_{\theta}(\bY)  \d\theta$ is the marginal likelihood. The posterior distribution of $\btheta$ is $\pexact(\btheta) = \pi(\btheta) p_{\theta}(\bY) / p(Y)$.

\section{Introduction \label{sec:intro}}

\section{Bayesian inference \label{sec:infer}}

We aim at performing Bayesian inference on the previously described model.
We set the following standard prior distribution on $\theta = (\alpha,\beta,\nu) = (\gamma,\nu)$: 

\begin{align}\label{prior}
\gamma = (\alpha, \beta)  & \sim \Ncal\left(\gamma_0, V_0 \right) \\
\nu & \sim \Dcal(e_0) \qquad \mbox{with }  e_0 = (e_{0, k})_{1 \leq k \leq K} \nonumber
\end{align}
where $\Dcal$ is the Dirichlet distribution.

In general for Bayesian inference, two strategies can be considered : either supplying a sample from the posterior distribution through a Monte Carlo method or approximating the posterior distribution in a given family of distribution as it is done for instance by the Variational Bayes estimation \citep[see][]{BeG03} 
or the Expectation Propagation  method \citep{Minka:2001}.

On the one hand, in the case of binary SBMs ($Y_i \in \{0,1\}$), the variational Bayes approximation  has proved its efficiency in terms of computational time and accuracy \citep{LBA12,LaR16}. However, its extension to the Poisson SBM with covariates is not straightforward and has not been solved yet. Note that in the frequentist context, a variational maximum likelihood  estimation of the parameters can be achieved \citep{MRV10}, but no uncertainty on the parameter estimates is provided. 
On the other hand, sampling methods such as Monte Carlo Markov Chain or Sequential Monte Carlo may be implemented but can be slow at exploring the posterior distribution due to the high dimension of the latent variables space \citep[see e.g.][]{NoS01}. 

{In this paper, we propose a two steps strategy. First, we derive an approximation of the posterior distribution of the parameters from a frequentist variational maximum likelihood estimate (see Subsection \ref{sec:VAR}). In a second step, we design an efficient Monte Carlo sampler taking advantage of the first approximation of the posterior distribution (see Subsection \ref{sec:SMC}). } 


\subsection{Derivation of an approximation for the posterior distribution \label{sec:VAR}}

\paragraph{Variational estimate of $\theta$.} 
Because the vector $Z = (Z_i)_{1 \leq i \leq n}$ of node memberships is unobserved, the likelihood of the data can not be easily evaluated. In a frequentist setting, the most popular approach resorts to the Expectation-Maximisation (EM) algorithm \citep{DLR77}, which requires the evaluation of some moments of the conditional distribution of the unobserved variables $Z_i$ given the observed $Y$. Unfortunately, this conditional distribution itself turns out to be intractable for SBMs \citep{NoS01}. For the unweighted SBM, \cite{DPR08} designed a variational version of EM \citep[VEM: see][for an introduction]{BKM17}. The VEM aims at maximizing a lower bound of the log-likelihood defined as
\begin{equation} \label{eq:J}
\begin{array}{ccl}
 J(Y; \theta, \qt)
 &:= & \log p_\theta(Y) - KL\left[\qt(Z) \;||\; p_\theta(Z \mid Y) \right] 
\\
 &= &  \Espt \log p_\theta(Y, Z)  + \mathcal{H}(\qt)
 \end{array}
\end{equation}
where $\qt(Z)$ is the approximation of $p_\theta(Z \mid Y)$ that minimizes the K\"ullback-Leibler (KL) divergence in a given family of distributions and $\Espt$ is the expectation over $Z$ according to $\qt$, $\mathcal{H}$ being the entropy. The approximate distribution $\qt(Z)$ is chosen among factorisable distributions, which results in a mean-field approximation \citep{WaJ08}:
\begin{equation}\label{eq:qt}
\qt(\bZ) = \qt(Z_1,\dots, Z_n ) =\prod_{i=1}^n \prod_{k=1}^K  \tautik^{\ind(Z_i=k)},
\end{equation}
where the variational parameter $\tautik$ is an approximation of the conditional classification probability $P(Z_i = k \mid Y)$. The lower bound given in \eqref{eq:J} becomes
\begin{align}
  J(Y; \theta, \qt) =& \sum_{i, k} \tautik \log \nu_k + \sum_{i<j} \sum_{k, \ell} \tautik \tautjl \log  f_{\Pcal}(Y_{ij};e^{\alphakl + \xij^\intercal \beta})  - \sum_{i, k} \tautik \log \tautik \nonumber  \\
  = & \sum_{i, k} \tautik \log \nu_k + \sum_{i<j} \sum_{k, \ell} \tautik \tautjl \left(-e^{\alphakl + \xij^\intercal \beta} + Y_{ij}(\alphakl + \xij^\intercal \beta) - \log Y_{ij} !\right) \label{eq: calcJ}\\
 & - \sum_{i, k} \tautik \log \tautik .  \nonumber
\end{align}
The \PSBMreg version of this algorithm has been introduced by \cite{MRV10}. It iterates until convergence the following  iteration $(t)$ 
\begin{description}
 \item[VE step:]
 $$
 \tauik^{(t)} 
 \propto \nuk^{(t)} \prod_{j \neq i} \prod_\ell \Pcal(\Yij; e^{\alphakl^{(t)} + \xij^\intercal \beta^{(t)}})^{\taujl^{(t)}}, 
\qquad \text{s.t. } \sum_k \tauik^{(t)} = 1; 
 $$
 \item[M step:]
 $$
 \theta^{(t+1)}
 = \arg\max_{\nu, \alpha, \beta} \sum_{i, k} \tauik^{(t)} \log \nu_k 
 + \sum_{i<j} \sum_{k, \ell} \tauik^{(t)} \taujl^{(t)} 
 \left((\alphakl + \xij^\intercal \beta) \Yij - e^{\alphakl + \xij^\intercal \beta} \right).
 $$
\end{description}

The \textbf{M step} actually consists in a weighted Poisson regression and can be obtained via gradient descent. The algorithm is implemented in the {\tt blockmodels} R package \citep{Leg16}. 
 The VEM algorithm results in a variational estimate $\thetat = ( \alphat, \betat,\nut) = (\gammat ,\nut) $, which we will use as a first guess for the posterior mean of the parameter. 


\paragraph{A proxy for the posterior distribution.}\label{sec:heuristic}
We now   design a proxy of the posterior distribution  $p_ {\bY} (\btheta)$ persuing  the variational strategy. A popular approximation of the posterior distribution  $ p_ {\bY} (\btheta) \propto \exp \left(\log \pi(\theta) + \log p_{\theta}(Y) \right)$  arises from the Laplace approximation, which results in  a Taylor expansion of the log-likelihood $ \log p_{\theta}(Y)$. This quantity being unavailable in our model, we propose to replace it  with its lower bound $J(Y; \theta, \qt)$  and perform a Taylor expansion of this quantity:  
\begin{align} \label{eq:Laplace}
 p_ {\bY} (\theta)
 & \propto \exp \left(\log \pi(\btheta) + \log    p_{\btheta}(Y) \right) \nonumber \\
 & \simeq \exp \left(\log \pi(\btheta) + J(Y; \btheta, \qt) \right) \\
 & \simeq \exp\left(\log \pi(\btheta) + J(Y; \widetilde{\btheta}, \qt) + \frac12 (\btheta-\widetilde{\btheta})^\intercal \left(\partial^2_{\theta^2} J(Y; \widetilde{\btheta}, \qt)\right) (\btheta-\widetilde{\btheta}) \right), \nonumber
\end{align}
where $\widetilde{\theta} = (\widetilde{\gamma},\widetilde{\nu})   :=\arg\max_\theta J(Y; \theta, \qt)$, is provided by the VEM algorithm. As shown in Appendix \ref{app:proxy}, the Hessian matrix $\partial^2_{\theta^2} J(Y; \theta, \qt)$ is made of two diagonal blocks corresponding to $\nu$ and $\gamma$, respectively. One may take advantage of this block-diagonal structure and define $\Vt_Y  := - (\partial^2_{\gamma^2} J(Y; \widetilde{\theta}, \pt))^{-1}$. Combining the Gaussian prior distribution on $\gamma$ defined in Equation  \eqref{prior},  Equation \eqref{eq:Laplace} suggests the following Gaussian proxy for the posterior of $\gamma$:
\begin{equation}\label{eq:post tilde gamma}
\papprox(\gamma) := \Ncal\left( \left(V_0 ^{-1} + \Vt_Y^{-1}\right)^{-1} \left(V_0 ^{-1} \gamma_0 +\Vt_Y^{-1} \widetilde{\gamma} \right), \left(V_0 ^{-1} + \Vt_Y^{-1}\right)^{-1} \right).
\end{equation} 
Regarding the vector of proportions $\nu$, we combine the Dirichlet prior distribution with the result of the VEM inference. Indeed, the VEM algorithm provides an estimate of the number of nodes belonging to each class $k$: $\widetilde{N}_k := \sum_i \tautik$. The conjugacy properties of the Dirichlet distribution suggest the following proxy for the posterior:
\begin{equation}\label{eq:post tilde nu}
\papprox(\nu) := \Dcal(e_0 + \et ), \qquad \text{where} \quad \et = (\widetilde{N}_k)_{1 \leq k \leq K}.
\end{equation} 
 Finally, using  $\qt(Z)$ as a proxy for $p_{\bY}(\bZ)$, we can combine equations \eqref{eq:qt}, \eqref{eq:post tilde gamma} and \eqref{eq:post tilde nu} to design a proxy $\papprox$ for  $p_{\bY}(\bZ,\theta)$: 
\begin{equation}\label{eq:papprox}
 \papprox(\bZ, \btheta) := \qt(\bZ) \papprox(\nu)  \papprox(\gamma). 
\end{equation} 
As a conclusion, $\papprox $ is a distribution combining the prior distribution and the data $\bY$.  The probabilistic dependence between the components of $\gamma$ are represented. However, $\papprox$ neglects the probabilistic dependence involving  $\bZ$.   The computational cost  of the computation  $ \papprox $  mainly reduces to  the implementation of a variational EM, which is known  to be  economical  from a computational point of view.  Besides,  $\papprox$ can be easily intensively simulated and   its density function   has an explicit expression. So, although $\papprox$ is not a satisfactory approximation of the posterior distribution,  we  claim that it can be used to drastically accelerate the posterior sampling of the true posterior distribution $\pexact$.

\subsection{Accelerated posterior sampling \label{sec:SMC}}
 
The main objective is to sample from the posterior distribution $ \pexact$.  A first  approach would consist in resorting to $\papprox$  as an  importance sampling (IS)  distribution.  However, this strategy is obviously naive  since there is no guarantee that the support of  $\papprox$ includes the support of the true distribution. 
As a consequence,   there is no hope to efficiently sample using 'one-step' IS. 
We propose to resort to an annealed importance sampling procedure \citep{Neal2001},   progressively shifting from the initial proposal $\papprox$ to the true posterior distribution $\pexact$.

An annealed importance sampling procedure consists in designing a sequence of distributions  $(\ph)_{h = 0\dots H}$ where $p_0$  is an easy simulated distribution  and $p_H$ is the distribution of interest,  in our case  $p_H = \pexact$. A classical choice for $(\ph)_{h = 0\dots H}$ proposed by \cite{Neal2001} is to consider 
$
 \ph(\bZ ,\btheta  )  \propto  \pi(\btheta)p_{\theta}(\bZ) \left(p_{\bZ,\btheta}(\bY)\right) ^{\rho_h} 
$ 
where $\rho_0 = 0$, $\rho_H = 1$, thus   moving from $\pi(\btheta) p_{\btheta}(\bZ)$ to the posterior $  \pexact(\btheta,\bZ)  \propto  \pi(\btheta)p_{\btheta}(\bZ)  p_{\bZ,\btheta}(\bY)$ by progressively integrating the data $\bY$ through the likelihood function. 
However, starting the annealing path from the prior distribution is far from efficient. We propose to take advantage of $\papprox$. 

More precisely,  we propose an alternative scheme moving smoothly from the approximate posterior distribution $\papprox$ to the true $\pexact$, setting the following  path: 
\begin{eqnarray}\label{eq:pih}
 \ph(\bZ ,\btheta ) & \propto & \papprox(\bZ,\btheta)^ {1-\rhoh}(\pexact( \bZ,\btheta)) ^{\rho_h}\nonumber\\
 & \propto &  \papprox(\bZ, \btheta) ^ {1-\rhoh}  (  p_{\btheta}(\bY,\bZ) \pi(\btheta) ) ^{\rho_h}.
\end{eqnarray}
where,  $\rho_0 = 0$, $\rho_H = 1$.  We claim that this scheme significantly reduces the computational time and is robust with respect to $ \papprox$ (see the numerical experiments in Section \ref{sec:illust}).\\ 
To sample from the sequence of distributions $(p_h)_{h = 1, \dots, H}$, we  resort to  the Sequential  Monte Carlo sampler  (SMC) proposed by \cite{DelMoral2006} where the annealing coefficients $(\rho_h)_{h=1,\dots, H}$ will be adjusted dynamically. 
  At iteration $h$, the SMC sampler involves three steps : moving the particles using a transition kernel, re-weighting the particles  in order to correct the discrepancy between the sampling distribution   and the distribution of interest at iteration $h$  (namely $\ph(\bZ ,\btheta ) $) and selecting the particles in order to reduce the variability of the importance sampling weights and avoid degeneracy. In practice, the particles will be resampled when the  Effective Sample Size ($ESS$)  decreases below a pre-specified rate. The algorithm is given below, the details being postponed to the Appendix   \ref{app:SMC}:

\label{page:algo}
 \vspace{1em}
\noindent\rule{\columnwidth}{0.4pt}

\noindent \textbf{Accelerated posterior sampling algorithm}
 \vspace{-0.5em}
 
\noindent\rule{\columnwidth}{0.4pt}

 \begin{enumerate}
 \item[] Set $(\tau_1,\tau_2) \in [0,1]^2$, $\rho_0 = 0$. 
 \item[0.] \emph{At iteration $0$} , sample $(\bthetazm)_{m = 1\dots M}$ from the approximate distribution $\papprox $ and set: 
 $$w_{0}^m = 1, \quad W_{0}^m = \frac{1}{M}, \quad r_{0}^m = \frac{p_{\bthetazm}(\bY)p_{\btheta_0^m}(\bZ_0^m) \pi(\btheta^m_{0})}{\qapprox(\bthetazm)}, \quad \quad  \quad  \forall m = 1\dots M.  $$
 \item[1.] \emph{At iteration $h$}: starting from $(\bthetahmoinsm, W_{h-1}^m,r_{h-1}^m)_{m = 1\dots M}$
 \begin{enumerate}
 \item Find (e.g. by binary search) $\rho_h$ such that: 
 $$ 
 \rho_ h = 1 \wedge \sup_{\rho}\left\{\rho > \rho_{h-1},cESS_{h-1}(\rho) \geq \tau_1 M \right\},
 $$ 
 where $$
cESS_{h-1}(\rho)\nonumber=   \frac{M \left(\sum_{m = 1}^M W_{h-1}^m (r_{h-1}^m)^{\rho -\rho_{h-1}}\right)^2}{\sum_{m = 1}^M W_{h-1}^m (r_{h-1}^m)^{2(\rho -\rho_{h-1})}}. 
$$

 \item $\forall m = 1\dots M $, compute $\whm = w^m_{h-1}\,\cdot \left(r_{h-1}^m\right)^{\rho_h - \rho_{h-1}}$ and $\Whm = {\whm}\left/{\sum_{m' = 1}^M w_h^{m'}}\right.$ 
 \item Compute 
 $$ESS_{h} = \frac{\left(\sum_{m = 1}^M\Whm\right)^2 }{\sum_{m = 1}^M (\Whm)^2} \in [1, M]$$ If $ESS_{h} < \tau_2\, M$, resample the particles 
 \begin{equation*}
 \begin{array}{ccl}
 (\bthetahmoinsm)' & \sim_{i.i.d} & \sum_{m = 1}^M \Whm \delta_{\{ \bthetahmoinsm\}}\\
 \bthetahmoinsm & \leftarrow & (\bthetahmoinsm)'\\
 w_{h}^m &  \leftarrow  & 1 \\
 W_{h}^m & \leftarrow  & 1/M
 \end{array}
 \quad \forall m = 1\dots M
 \end{equation*}
 \item $\forall m = 1\dots M$, : propagate the particle $(\bthetahm) \sim K_h( \cdot \mid \bthetahmoinsm) $ where $K_h$ is a MCMC kernel with $\ph(\bZ,\btheta) $ 
as an invariant distribution and compute:
 $$ r^m_{h} = \frac{p_{\btheta_h^m}(\bY,\bZ_h^m) \pi(\btheta_h^m)}{\qapprox(\bthetahm)} $$ 
 \end{enumerate}
 \item[2.] If $\rho_h = 1$, stop. If $\rho_h < 1$ return to $1$. 
\end{enumerate}

\vspace{-0.5em}
\noindent\rule{\columnwidth}{0.4pt}

The statistical properties of $\sum_{m=1}^ M W^H_m \phi(\theta^{H}_m)$  as an estimator of $ \Esp[\phi(\theta) \mid \bY]$  are    studied in \cite{Doucet2009} (and references therein).  First of all, $\sum_{m=1}^ M W^H_m \phi(\theta^{H}_m)$ is known to be strongly convergent. Moreover, following \cite{DelMoral2006}, a Central Limit Theorem can be obtained.  Besides, in addition to these asymptotic properties,  it is possible to control the mean-square error of  the estimator for a given  number of particles $M$, provided additional assumptions on $\phi$.   Results of convergence were also provided by \cite{delmoral2012_conv} for adaptive sequential Monte Carlo algorithms.

\subsection{Posterior inference}\label{sec:post}

\paragraph{Estimation of the marginal likelihood.}\label{subsec: log marg}

With respect to MCMC strategies, Annealing Importance Sampling and SMC have the great advantage to supply good estimators of the marginal likelihood, which is the critical quantity when model comparison or model selection is at stake. Following  \cite{DelMoral2006},  
\begin{equation}\label{eq:marg}
 \widehat{m}_{\bY} =  \prod_{h = 1}^H \sum_{m = 1}^M \Whm \left[r(\bthetahmoins)\right]^{\rho_h-\rho_{h-1}}
\end{equation}
is a consistent estimator of $p(\bY)$.  Details are provided in the Appendix \ref{app:SMC}. 

\paragraph{Model selection and averaging.}
The algorithm introduced in Section \ref{sec:SMC} is defined for a fixed number of groups $K$. As a consequence Equation  \eqref{eq:marg}  provides an  estimate  of $p(Y \mid K)$. In most practical cases, the number of groups $K$ is actually unknown but, for a given prior $\pi(K)$, it can then be estimated by maximizing the posterior distribution $\widehat{p}_{\bY}(K) \propto \pi(K) \widehat{p}(Y \mid K)$. This requires to run the algorithm for a series of value of $K$.\\
Similarly, the proposed algorithm provides samples from the posterior conditional distribution of the model's parameter $p(\theta \mid Y, K)$. To this respect the weighted sample $\{(\theta^m, W^m)\}_m$ should rather be denoted $\{(\theta^{m, K}, W^{m, K})\}_m$. Picking a specific value for $K$ is not required to make inference on the parameters that do not depend on it, such as the regression coefficients $\beta$. Indeed a sample from their marginal posterior distribution can be obtained by model averaging. It suffices to gather the samples $\{(\beta^{m, K}, W^{m, K})\}_{m}$ obtained with each $K$ into the larger sample $\{(\beta^{m, K}, \overline{W}^{m, K})\}_{m, K}$ and to rescale the weights $\overline{W}^{m, K} := \widehat{p}_{\bY}(K) W^{m ,K}$.

\paragraph{Residual structure.}
Network analysis in presence of covariates raises two typical questions. The first one is the actual effect of each of these covariates on the structure of the network and the second one is the existence of some residual structure in the network, once accounted for the effect of the covariates. 
The  inference on the $\beta$ provides answers to the first one. As for the second one, we use the residual representation introduced by \cite{LRO18}, which we adapt by rephrasing the \SBMreg model in the following way: each node $i$ is associated with a uniform draw $U_i$ over the unit interval and the interaction $Y_{ij}$ is then drawn conditionally on $(U_i, U_j)$ as a Poisson variable with mean $\exp(\phi(U_i, U_j) + \xij^\trans \beta)$. For a $K$-block SBM, $\phi : [0, 1]^2 \mapsto \mathbb{R}$ is a rectangular block-wise constant function, with block widths $(\nu_k)$ and block heights $(\alpha_{k\ell})$ (see Figure \ref{fig:graphon}, top left). Using this representation, each particle $\theta^{m, K}$ corresponds to a function $\phi^{m, K}$. A map of the residual structure (conditional on $K$) can then be obtained as $\widehat{\phi}^K(u, v) = \sum_m W^{m, K} \phi^{m, K}(u, v)$. An unconditional estimate $\widehat{\phi}$ can also be derived by averaging over $K$. 

The residual structure can be further investigated at the node level via the latent coordinates $U_i$. Indeed, $U_i$ is independent of $Y$ conditionally on $\theta$ and $Z_i$ and is uniformly distributed over $(\nu^+_{Z_i-1}; \nu^+_{Z_i})$, where $\nu^+_0 = 0$ and $\nu^+_k = \sum_{\ell = 1}^k \nu_k$ for $1 \leq k \leq K$. The posterior mean of each $U_i$ can hence be estimated  by averaging over all the particles $\theta^{m}$ and $Z^{m}$, conditionally or unconditionally on $K$.

\section{Illustrations \label{sec:illust}}

We now illustrate the relevance and the efficiency of our inference method, first  on  datasets simulated from the model (Subsection \ref{subsec:simu}) and then on two datasets  issued from the ecological field (Subsection \ref{subsec:data}). 

\subsection{Simulation study \label{subsec:simu}}

We first present a  simulation study    assessing the fact that our    strategy  combining   the design of an ad-hoc proxy $\papprox$ for the posterior  and its use to sample from  the true  posterior distribution    drastically decreases the computational time with respect to a classical annealing-scheme  (starting from the prior distribution) or, equivalently, that    $\papprox$  can be "corrected" into the true posterior  distribution at a low computational cost. 

Note that this two-steps strategy has been tested on other statistical  models in a previous working paper \cite{donnet:hal-01566898}
where the robustness of the sampling strategy with respect to the mis-specification of $\papprox$ is also tested.

\paragraph{Simulation design.} 
We simulate $S = 100$ networks with $n = 40 $ nodes according to the  Poisson SMB-reg model  with $K =  2$ groups and $p = 4$ covariates, the covariates matrix $X$ being fixed to an arbitrarily chosen value.   The $S$ replicates are simulated using the following scheme.  
The parameters $\nu^{(s)}$ and  $\gamma^{(s)}$ are generated from the prior distribution defined in Equation \eqref{prior} with  the following hyperparameters: 
 \begin{equation}\label{eq:paramsim}
\gamma_0  =  (1, \; 0, \; 3, \; 1.1, \; 2.2, \; 0.1, \; -0.3), \qquad 
V_0 =  0.1 \cdot {\bf I}_7, \qquad 
e_0 =  (3, \; 3),
\end{equation} 
where ${\bf I}_d$ is the identity matrix of size $d\times d$. For each simulated parameter $\btheta^{(s)}$,   a dataset $\bY^{(s)}$ is simulated according to the Poisson SMB-reg model defined in Equations \eqref{eq:Z} and  \eqref{eq:Y}. 
The resulting weighted networks  are such that the $\left(Y^{(s)}_{ij}\right)_{i,j,s}$ belong to $\{0,\cdots, 187\}$ with  $\mathbb{E}[Y_{ij}] \approx 10$ and  $\Var(Y_{ij}) \approx 13^2$. 
We first consider that the number of groups $K$ is known and we focus on the posterior distribution of the parameters $\theta$. 

For each dataset $Y^{(s)}$, we aim at sampling the posterior distribution corresponding to the informative prior distribution  $\pi(\cdot)$ defined in equation \eqref{prior} with hyperparameters given in  \eqref{eq:paramsim}.  First of all, we derive the proxy of the posterior distribution using the approach presented in Section \ref{sec:VAR}  resulting  into $\papproxs$ (\textsf{Approx} in the legend). Then we  use   the   sampler presented in Section \ref{sec:SMC}  to  get a  sample  from the true posterior distribution $p_{\bY^{(s)}}$   (this method is referred as the \textsf{SMC from approx} strategy in the plots).  
Finally, we compare  our strategy with  a   SMC applied to a standard  annealing scheme starting from the prior distribution (referred as \textsf{SMC from prior}). 
The strategies are  compared in terms of computation time and accuracy. 

The two SMC algorithms are  run with $ M  = 2000$ particles, $\tau_1 = 0.9$, and $\tau_2 = 0.8$.  All the codes are written in \textsf{R}. The variational estimation  is  obtained from  the \textsf{R}-package \textsf{blockmodels} \citep{Leg16}. 

\paragraph{Computational time.} For each dataset, we first compare the number of iterations in the SMC procedure required to go either from  the prior distribution $\pi$ to $p_{Y^{(s)}}$  or from $\papproxs$ to  $p_{Y^{(s)}}$.  Remember, that the sequence $(\rho_h)_{h=1,\dots, H}$ is not a tuned parameter of the algorithm but is self-tuned in an adaptive way. As a consequence, the number of iterations (i.e. the  length of the sequence $(\rho_h)_{h=1,\dots, H}$)  is a first rough indicator of how good  the proxy  $\papprox$ is as an approximation of the true posterior distribution $p_{\bY}$. 

\begin{figure}[ht]
 \begin{center}
  \includegraphics[width=0.7\textwidth]{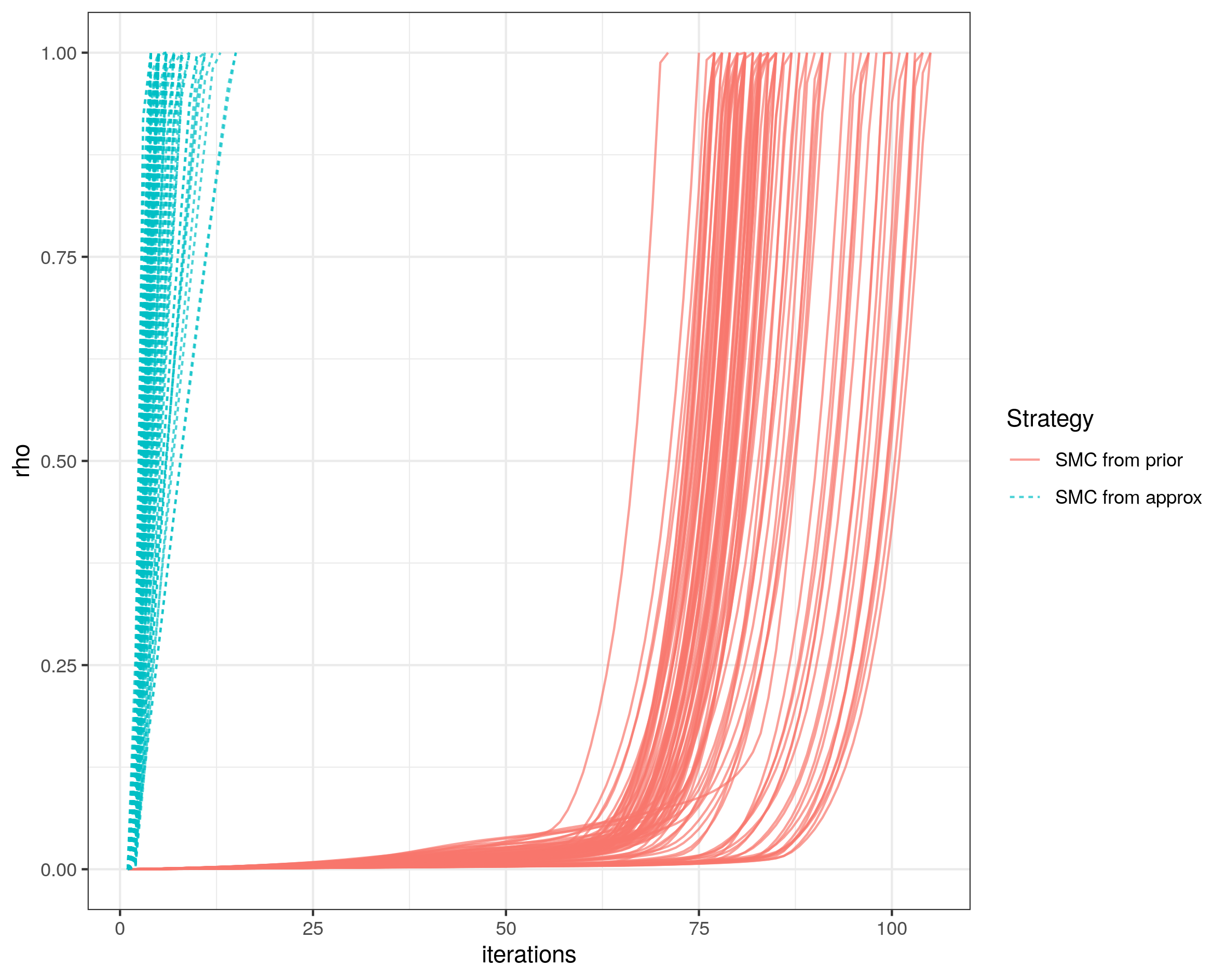}
 \end{center}
 \caption{Simulated dataset :  sequences of $(\rho_h)$ along iterations for each dataset and each strategy. 
 \label{fig:traj_rho}
 }
\end{figure}

In  average,  starting from  $\widetilde{p}_{Y^{(s)}}$ results into a number of iterations fifteen  times smaller, going from  an averaged number of   $85$ iterations in the \textsf{SMC from prior} strategy to around  $6$ iterations in the \textsf{SMC from approx} strategy. The sequences $(\rho_h)$    for each dataset $s$ and each strategy are plotted in Figure \ref{fig:traj_rho}. As expected, starting from $\widetilde{p}_{Y^{(s)}}$ induces a clear decrease of the number of iterations. 

In terms of computational time, we performed the experiments on a  \verb+Intel®  Xeon(R)+    CPU  E5-1650 v3 @ 3.50GHz x12+ using $6$ cores. For such a network, the \textsf{R}-package \textsf{blockmodels}  supplies the variational estimation of interest in less than $1$ minute  (performing at the same time the selection of the number of blocks $K$).  The \textsf{SMC from approx} strategy terminates in  (in average)  32 seconds. The computational time is about fifteen times longer for \textsf{SMC from prior}.

\begin{figure}[ht]
 \begin{center}
  \includegraphics[width=\textwidth]{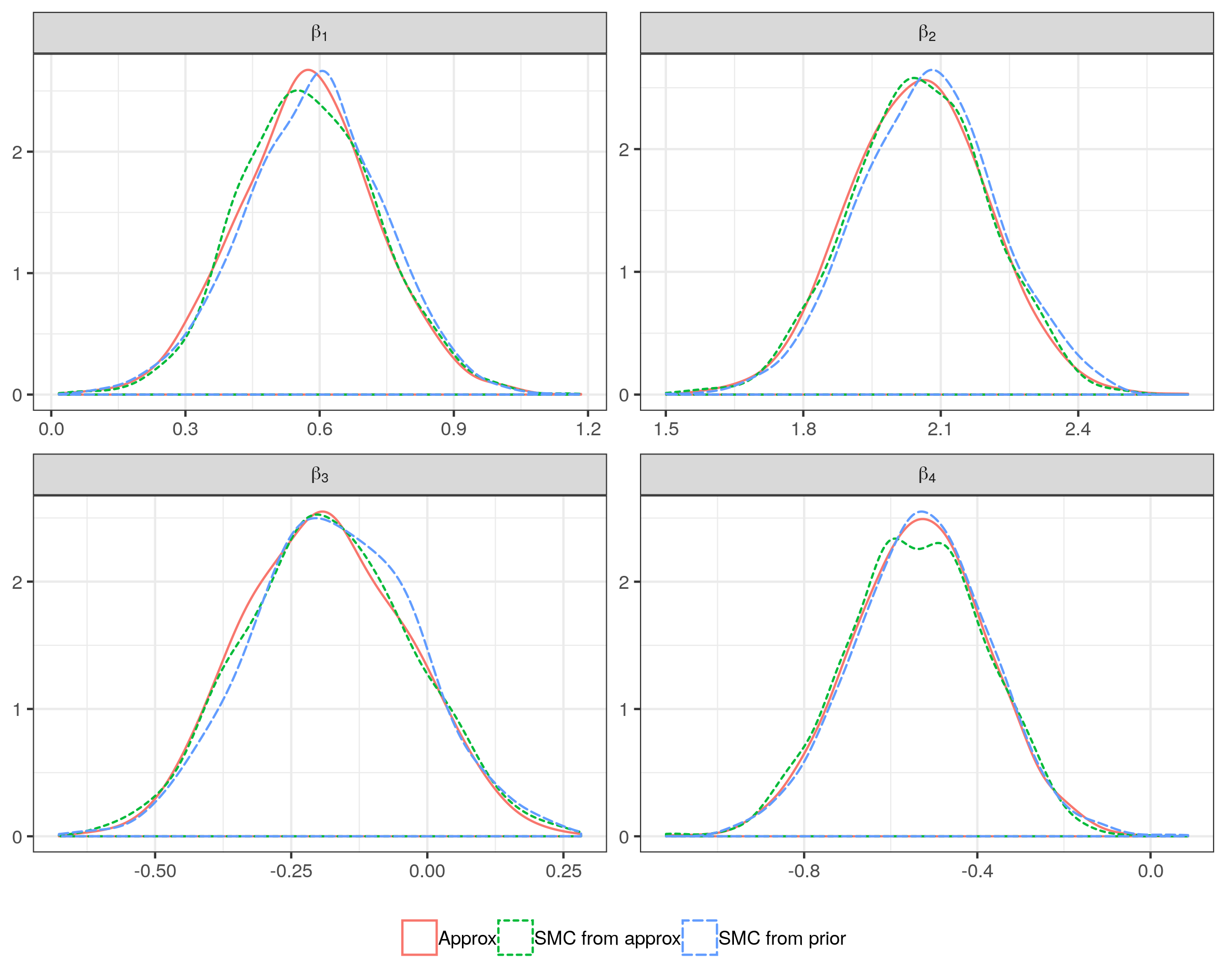}
 \end{center}
 \caption{Simulated dataset. Posterior distribution of $(\beta_k)_{k=1,\dots,4}$ given by the Standard SMC (Stand. SMC), the variational estimation (VEM), the VEM combined  with the prior distribution   (VEM and prior) and the SMC starting from $\papproxs$ (VEM +  SMC)  on a arbitrarily chosen simulated dataset. 
 \label{fig:post_beta_simu45}
 }
\end{figure}

\paragraph{Posterior distributions of the parameters.}
We now compare the posterior distributions of the parameters.  Figure \ref{fig:post_beta_simu45} supplies  an example (on one given simulated dataset) of the posterior distributions of ($\beta_1, \beta_2, \beta_3,\beta_4$) supplied  by the  three strategies. On this dataset (as well as on all other simulated dataset), the posterior distribution supplied  by the \textsf{SMC  from prior} and the \textsf{SMC from approx}  are similar. The robustness of our strategy was already illustrated in a previous work on a wider variety of models \citep{donnet:hal-01566898}.   Besides,  Figure \ref{fig:post_beta_simu45} illustrates the fact that  $\widetilde{p}_{Y^{(s)}}(\beta_k)$   already supplies a good approximation of the true posterior marginal distribution  $p_{Y^{(s)}}(\beta_k )$ since the red  plain line coincides the other curves.
\begin{figure}[ht]
 \begin{center}
  \includegraphics[width=0.5 \textwidth]{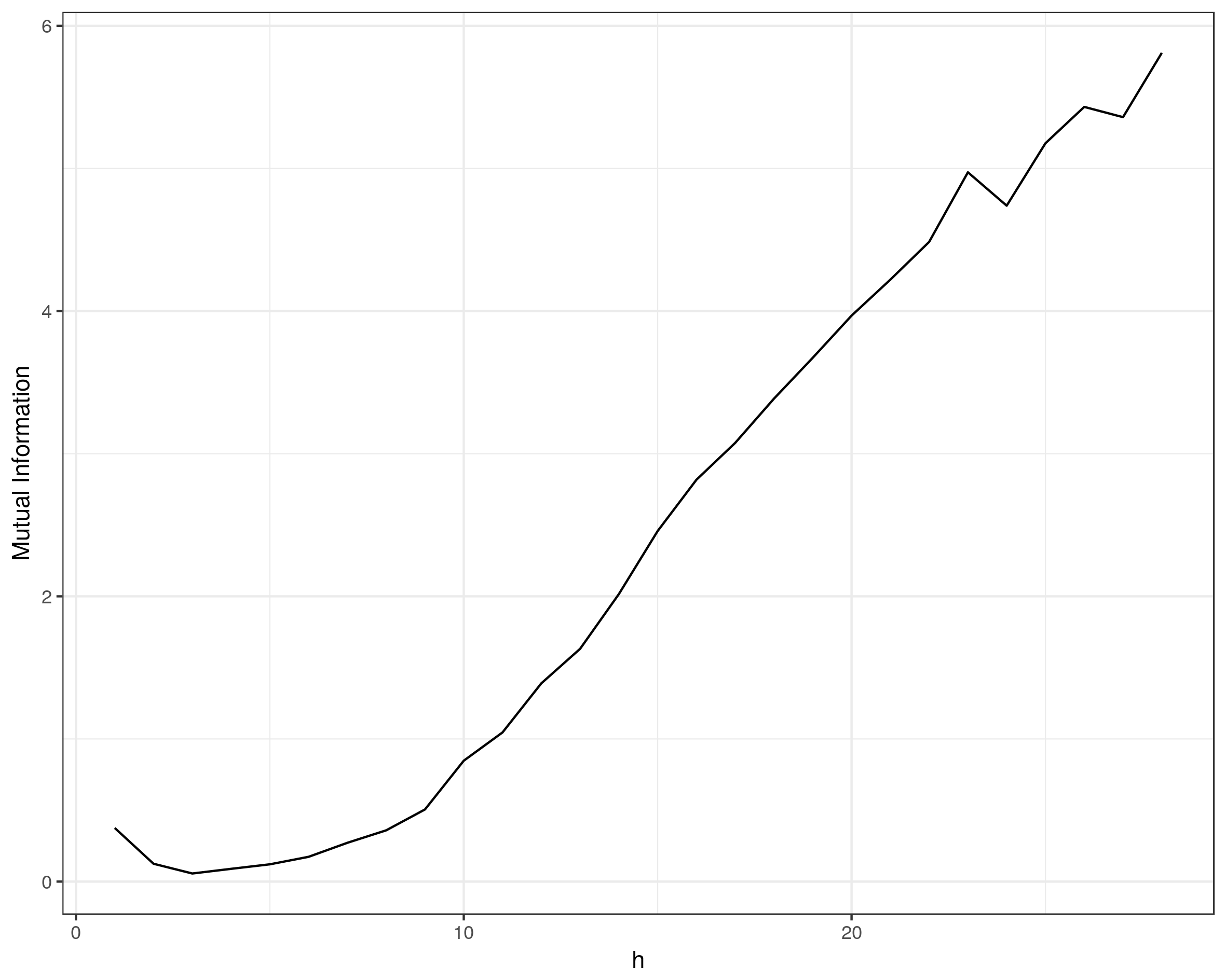}
 \end{center}
 \caption{\emph{Simulation  study}. Mutual interaction $\text{MI}_h(\bZ)$ along iterations 
 \label{fig:MI_simu45}
 }
\end{figure}
However, when focusing on the joint distributions,  we know that, by construction,  $\papproxs$ neglects    the probabilistic dependencies between   $\bZ$ and $\theta$ and also between the $(Z_i)$'s. The iterations required by the SMC algorithm are used to learn these dependencies.  In order to illustrate this phenomena, we monitor the Mutual Information  (MI) of the $\bZ$  along the iterations $h$:
\begin{equation}\label{eq:MIh}
\text{MI}_h(\bZ) = KL\left[p_h(\bZ)  ; \prod_{i=1}^n p_h(Z_i)\right]. 
\end{equation}
 $\text{MI}_h(\bZ)$ can not be computed exactly so we consider its empirical estimator  $\widehat{\text{MI}}_{h,M}(\bZ)$.   $\widehat{\text{MI}}_{h,M}(\bZ)$ is plotted in Figure  \ref{fig:MI_simu45}  for one arbitrarily chosen dataset.  As expected, this quantity increases along the iterations, confirming the fact that the sequential sampler learns the   dependence structure neglected  in $\papproxs$.

\paragraph{Validation.}
In order to validate our inference algorithm, we use the simulation-based validation tools proposed by \citet{donnet:hal-01566898} and \cite{Talts2018}. 
More precisely,  let us recall the fact  that,  if 
\begin{align}\label{eq:model}
 \theta  & \sim \pi(\cdot), \nonumber \\
Y \mid    \theta & \sim \ell( \cdot  \mid \theta ), \\
 \theta^{m}  \mid Y & \sim_{i.i.d.} p(\cdot \mid Y), & & m=1,\dots, M, \nonumber
\end{align}
then, for any  real-valued function $\Phi(\theta)$, we have
\begin{equation} \label{eq:crit-theo}
 U(\btheta, \bY,\Phi,(\theta^{m})_{m = 1,\dots, M}) =  \sum_{m=1}^M \ind_{\Phi(\theta^{m}) < \Phi(\theta ^{})}   \sim  \mathcal{U}_{\{1, \dots, M\}}. 
\end{equation}
As a consequence, we propose the following validation strategy.  Let $\Phi$ be fixed.  For $s= 1\dots, 100$,  let $(\theta^{(s)},  Y^{(s)})$  be a realisation  of distributions  \eqref{eq:model} (with hyperparameters given in \eqref{eq:paramsim}). For any $s$,   let  $(\widetilde \theta^{(s)m})_{m=1,\dots, M}$  be a sample from $\widetilde{p}_{Y^{(s)}}$ and let $(\theta^{(s)m}))_{m=1,\dots, M}$ denote the sample obtained from the \textsf{SMC from approx} procedure;  $(\overline \theta^{(s)m}))_{m=1,\dots, M}$ is a sample  obtained with \textsf{SMC from prior}.  For any  $s=1\dots, 100$,  we compute: 
 \begin{equation*}
\widetilde{U}^{(s)}(\Phi) =  \sum_{m=1}^M \ind_{\Phi(\widetilde \theta^{(s)m}) < \Phi(\theta ^{(s)})}  \\
\end{equation*}
and $U^{(s)}(\Phi)$ and $\overline{U}^{(s)}(\Phi)$ for $\theta^{(s)m}$ and $\overline \theta^{(s)m}$, respectively,
and  compare the empirical distribution of   $(\widetilde{U}^{(s)}(\Phi)) _{s=1\dots S}$ ,  $(U^{(s)}(\Phi)) _{s=1\dots S}$  and $(\overline{U}^{(s)}(\Phi)) _{s=1\dots S}$ to the uniform one  $\mathcal{U}_{\{1, \dots, M\}}$.  
We apply our procedure for eleven different functions $\Phi$  all invariant under label switching  (the functions are provided in the Appendix  \ref{sec append phi}), .

\begin{figure}[ht]
 \begin{center}
  \begin{tabular}{cc}
  \includegraphics[width=0.5 \textwidth]{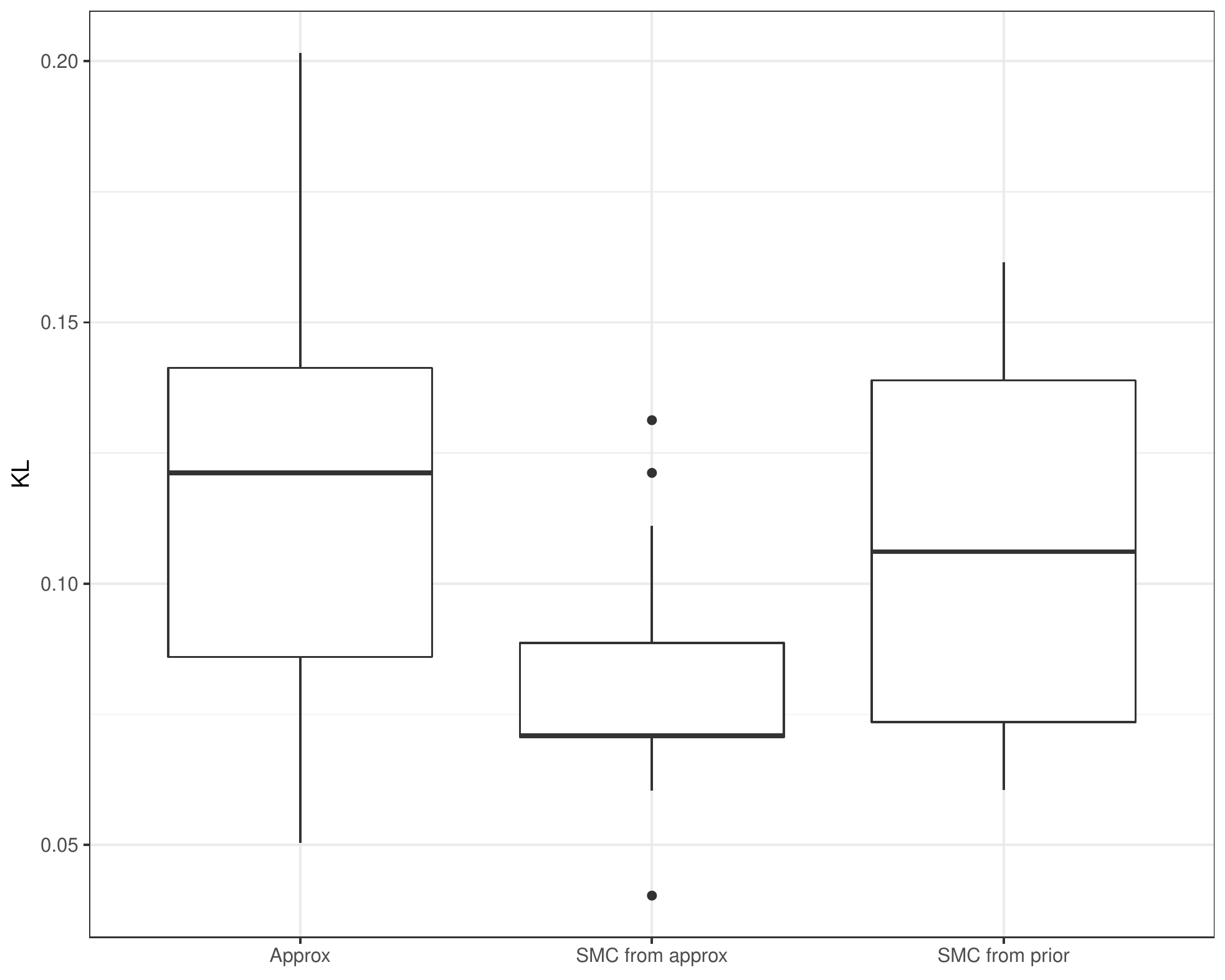} &
  \includegraphics[width=0.5 \textwidth]{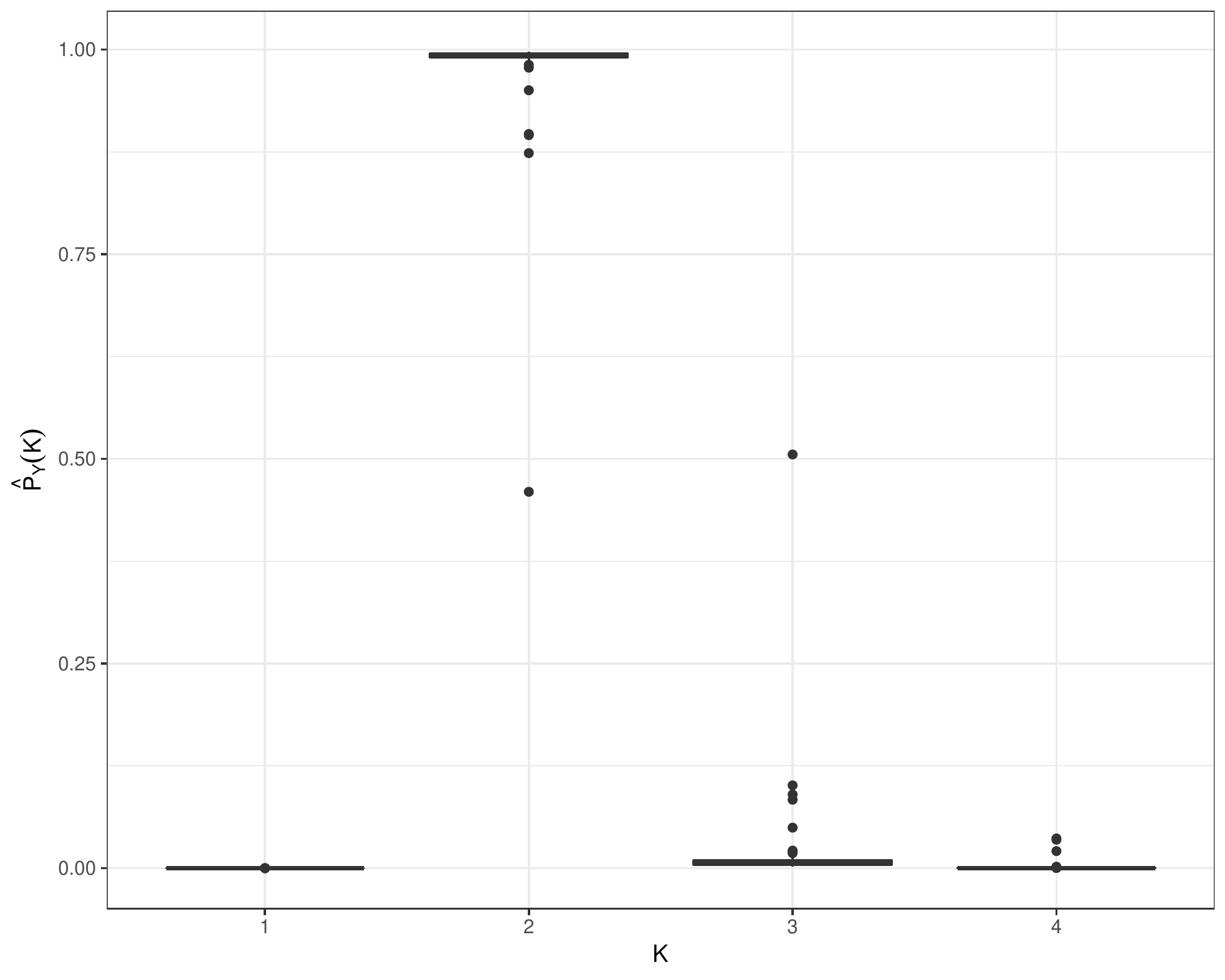}
  \end{tabular}
 \end{center}
 \caption{\emph{Simulation  study}.  Left: Boxplots of the KL divergence between the uniform distribution and   $(\widetilde{U}^{(s)}(\Phi))_{s=\dots S}$,  $(U^{(s)}(\Phi))_{s=1\dots S}$  and $(\overline{U}^{(s)}(\Phi))_{s=\dots S}$ respectively,   for the eleven functions  $\Phi$. Right: Boxplots of the $\hat{P}_{Y^{(s)}}(K))_{s=1,\dots,100}$ obtained with our method}
\label{fig:KL}
\end{figure}

In Figure \ref{fig:ecdf U}, we plot the empirical cumulative distribution function (ecdf)  of $(\widetilde{U}^{(s)}(\Phi))_{s=\dots S}$, $(U^{(s)}(\Phi))_{s=1\dots S}$ and  $(\overline{U}^{(s)}(\Phi))_{s=\dots S}$ for $\Phi$ equal to: 
\begin{equation*}
\begin{array}{lcllcl}
\Phi_4(\theta)  &=&  \beta_2, \quad  & \Phi_8(\theta)  &=& \sum_{k,\ell} \alpha_{k\ell} \\
\Phi_{10}(\theta)  &=& \sum_{k,\ell} \alpha_{k\ell}  + \beta_3, \quad  & \Phi_{11}(\theta)  &=&\sum_{k,\ell} \alpha_{k\ell}  + \sum_{r=1}^4 \beta_r + |\pi_1 - \pi_2|  
 \end{array}. 
\end{equation*}
We observe that for $\Phi_4$,  $\widetilde{p}_{Y^{(s)}}$ already supplies a good approximation of the posterior distribution. However, for  the other functions, the ecdf of  $(\widetilde{U}^{(s)}(\Phi))_{s=\dots S}$ is far from the targeted uniform distribution, probably because of its poor evaluation of the posterior dependency between the parameters.  The ecdf of $({U}^{(s)}(\Phi))_{s=\dots S}$ is much more similar to the uniform one.  Note that, as can be observed for $\Phi_8$,  \textsf{SMC from approx} $\widetilde{p}_{Y^{(s)}}$ performs better than \textsf{SMC from prior}, meaning that the posterior distribution is better explored when using a first approximation of the posterior distribution as $\widetilde{p}_{Y^{(s)}}$.  This phenomena is confirmed  in Figure \ref{fig:KL} (left panel) where we display  the boxplots of the KL divergence between the uniform distribution and   $(\widetilde{U}^{(s)}(\Phi))_{s=\dots S}$,  $(U^{(s)}(\Phi))_{s=1\dots S}$  and $(\overline{U}^{(s)}(\Phi))_{s=\dots S}$ respectively,   for the eleven functions  $\Phi$.    \textsf{SMC from approx} performs better  (i.e. KL smaller) than  not only $\widetilde{p}_{Y^{(s)}}$ but also \textsf{SMC from prior}. 

\begin{figure}[ht]
 \begin{center}
  \includegraphics[width= \textwidth]{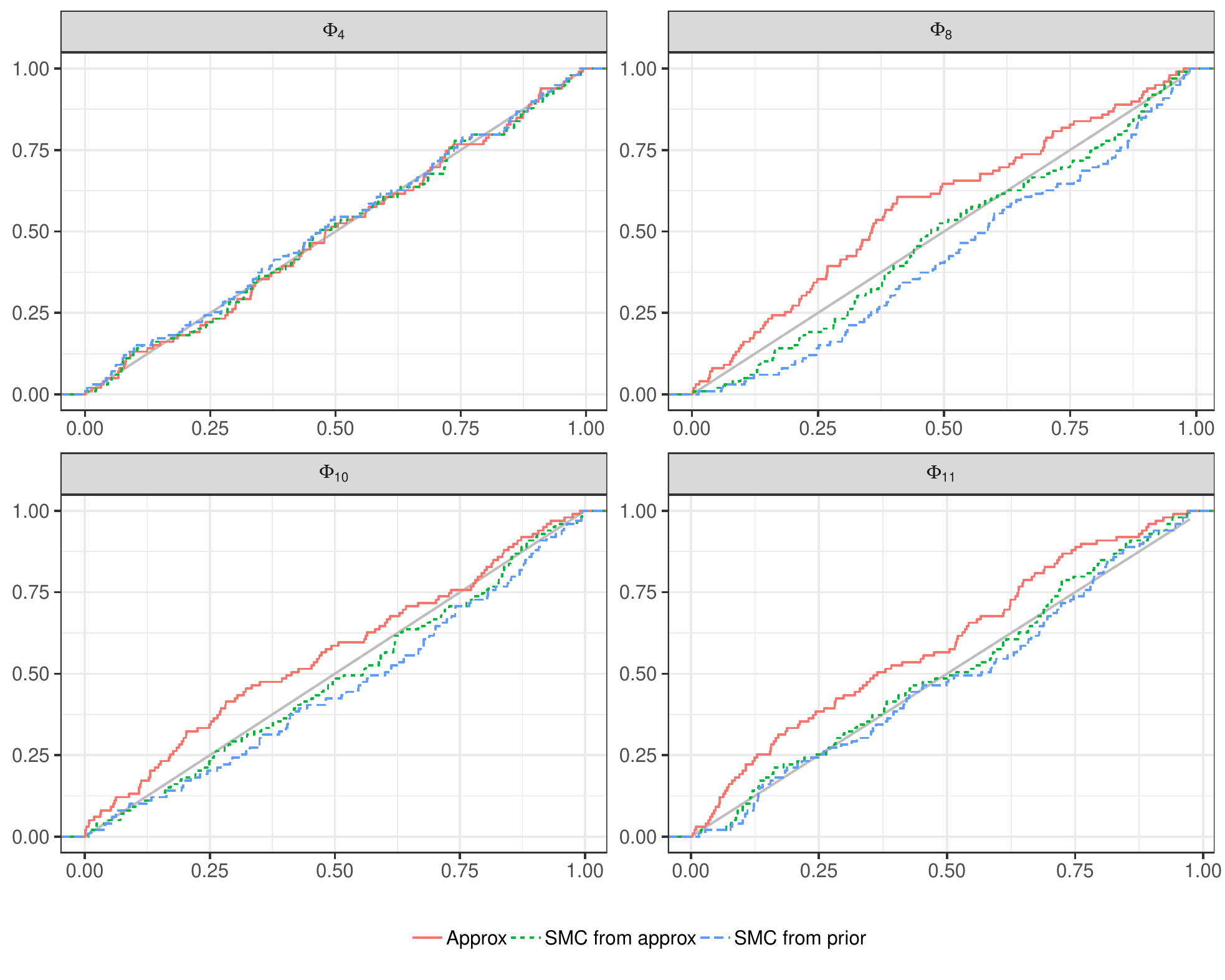} 
 \end{center}
 \caption{\emph{Simulation  study}.  Ecdf of $(\widetilde{U}^{(s)}(\Phi))_{s=\dots S}$ and $(U^{(s)}(\Phi))_{s=1\dots S}$  for $\Phi_{4},\Phi_{8},\Phi_{10},\Phi_{11}$ }
\label{fig:ecdf U}
\end{figure}

\paragraph{Selection of $K$} We can also perform model selection by running the algorithm for several values of $K$.  One of the advantage of the SMC algorithm is that is provides a Monte Carlo approximation of the marginal likelihood as exposed in Section \ref{subsec: log marg}.  The boxplots corresponding to $(\hat{P}_{Y^{(s)}}(K))_{s=1,\dots,100}$ are displayed in the right panel of Figure \ref{fig:KL}. We observe that the sampled posterior highly concentrates around the true value $K=2$.

\subsection{Application on ecological datasets \label{subsec:data}}

\newcommand{\figdir}{Figs}

\begin{figure}[ht]
 \begin{center}
  \begin{tabular}{ccc}
   \includegraphics[width=.3\textwidth]{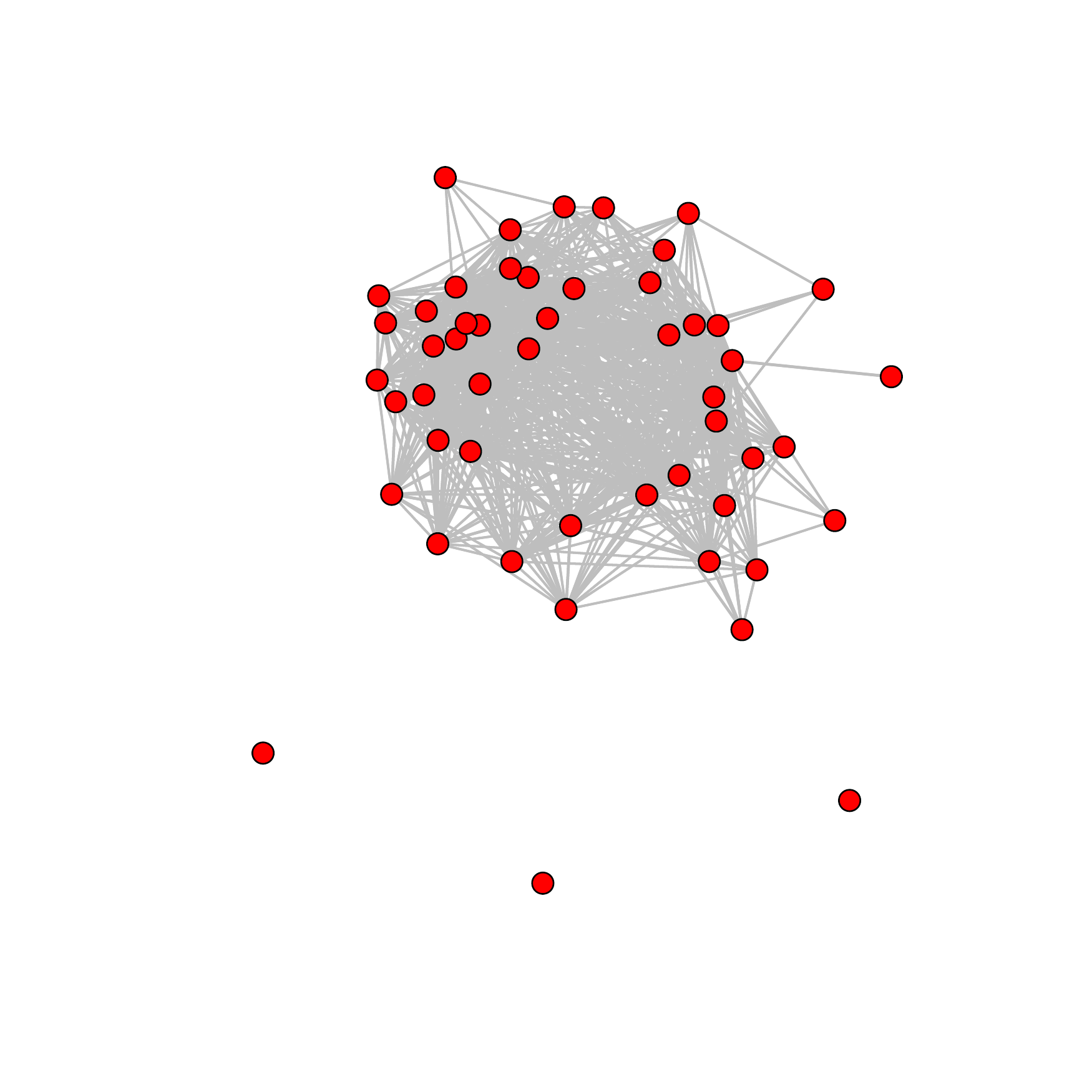} &
   \includegraphics[width=.3\textwidth]{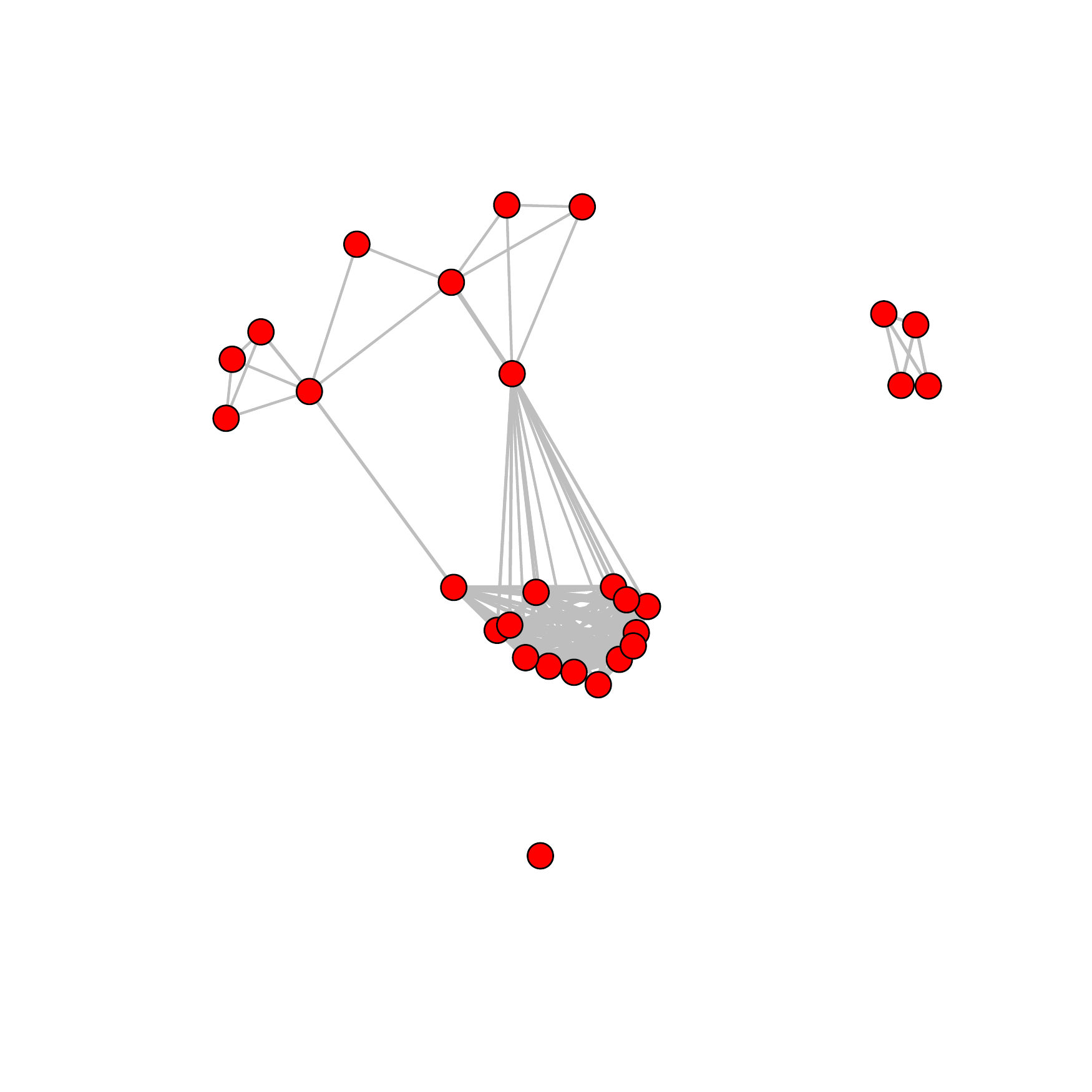} &
   \includegraphics[width=.3\textwidth]{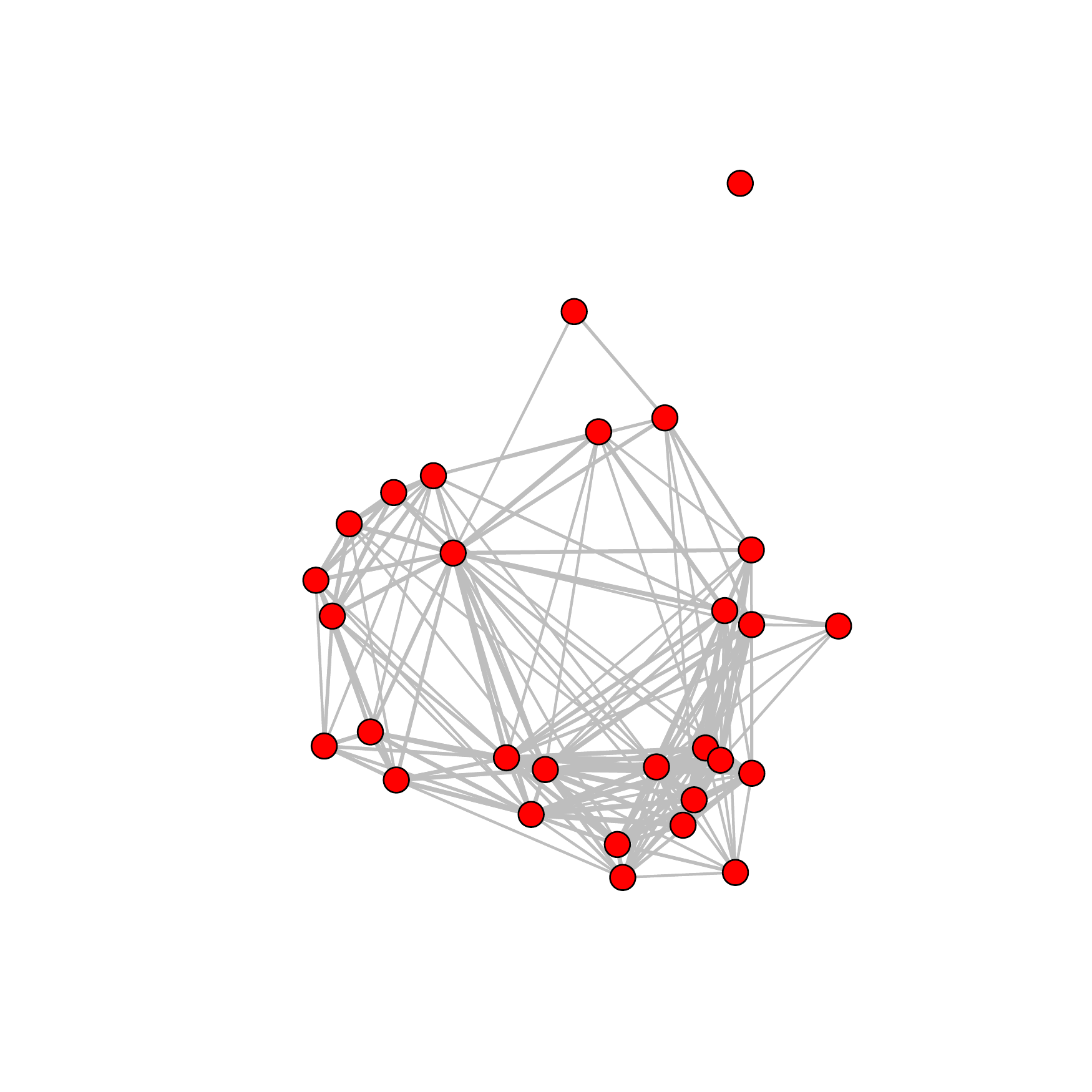}
  \end{tabular}
 \end{center}
 \caption{Observed weighted networks for trees (left), zebras (center) and onager (right).
 \label{fig:net}}
\end{figure}

\subsubsection*{Tree species parasitic network}

We first apply the proposed methodology to the tree network introduced by \cite{VPD08}. The data consists of a set of $n=51$ tree species. For each pair $(i, j)$ of species, the number $Y_{ij}$ of shared fungal parasites (that is: parasites that can be hosted by both species) was recorded. For each pair of species, three distances were also measured, namely the taxonomic ($x^1$), the geographic ($x^2$) and the genetic distance ($x^3$). 
The aim of the study was to exhibit clusters of species in the parasitic network, but also to see if the number of shared parasites depends on the similarity between species. If the latter holds, ecologists are primarily interested in defining species clusters that can not be mainly predicted by the distances between them \citep[see][which include the data in the supplementary material]{MRV10}.

Figure \ref{fig:tree} displays the results of this analysis. The number of clusters with highest posterior probability is $5$, whereas pseudo ICL criterion selects $4$ clusters. The second plot shows that it takes about 25 steps to sample from the posterior. 

Regarding the parameter inference, we first observe that the approximate posteriors of the regression coefficients $\beta_j$ (in blue) are quite close from their true posteriors. This may seem surprising at first glance that it takes more than 25 steps to make such a small shift. However, the posterior distributions of the regression coefficients $\beta_j$ are only marginals. Table \ref{tab:tree} provides the (approximate) posterior correlation between these coefficients and shows that $\papprox$  tends to under-estimate these correlations, which is consistent with the results from the simulation study. 


We also study the evolution of the distribution of the latent variables $Z_i$ along the sampling path. We remind that the variational approximation consists in assuming that they are conditionally independent, whereas they are not. Therefore, the proposed SMC algorithm should retrieve this dependency.  
{To observe this behavior, once again,  we monitor the estimated mutual information along the iterations  $\widehat{\text{MI}}_{h,M}(\bZ)$.} The top right panel of Figure \ref{fig:tree} shows that the estimated $\widehat{\text{MI}}_{h,M}(\bZ)$ does increase along the steps. This indicates that a substantial part of the sampling effort is dedicated to the reconstruction the conditional dependency structure of the $Z_i$, which is also consistent with the results from the simulation study.  

According to the posterior distribution of $\beta_3$, the effect of the contribution of the genetic distance is questionable. For all parameters, averaging the posterior of $K$ tends to smooth the posterior but does not have a strong effect on the inference. 
The model choice is balanced: 
the models with higher probability are ($x^1$) and ($x^1, x^2$), with respective probability $52.1\%$ and $46.8\%$. 
The models based on all other combination of variables have a posterior probability smaller than $10^{-2}$, which confirms the absence of effect of the genetic distance or its redundancy with respect to the other distances.

All these results are consistent with the conclusions of \cite{MRV10}: the distances between the species concur to structure the parasitic network (although the genetic distance does not bring a significant additional information). Still, these two distances are not sufficient to predict the whole topology of the network and a residual structure remains.

\begin{figure}[ht]
 \begin{center}
  \begin{tabular}{ccc}
   $p(K \mid Y)$ & $\rho_h$ & $\widehat{MI}_h$ \\
   \begin{tabular}{c} \includegraphics[trim=10 10 10 10, width=.3\textwidth, height=.15\textheight]{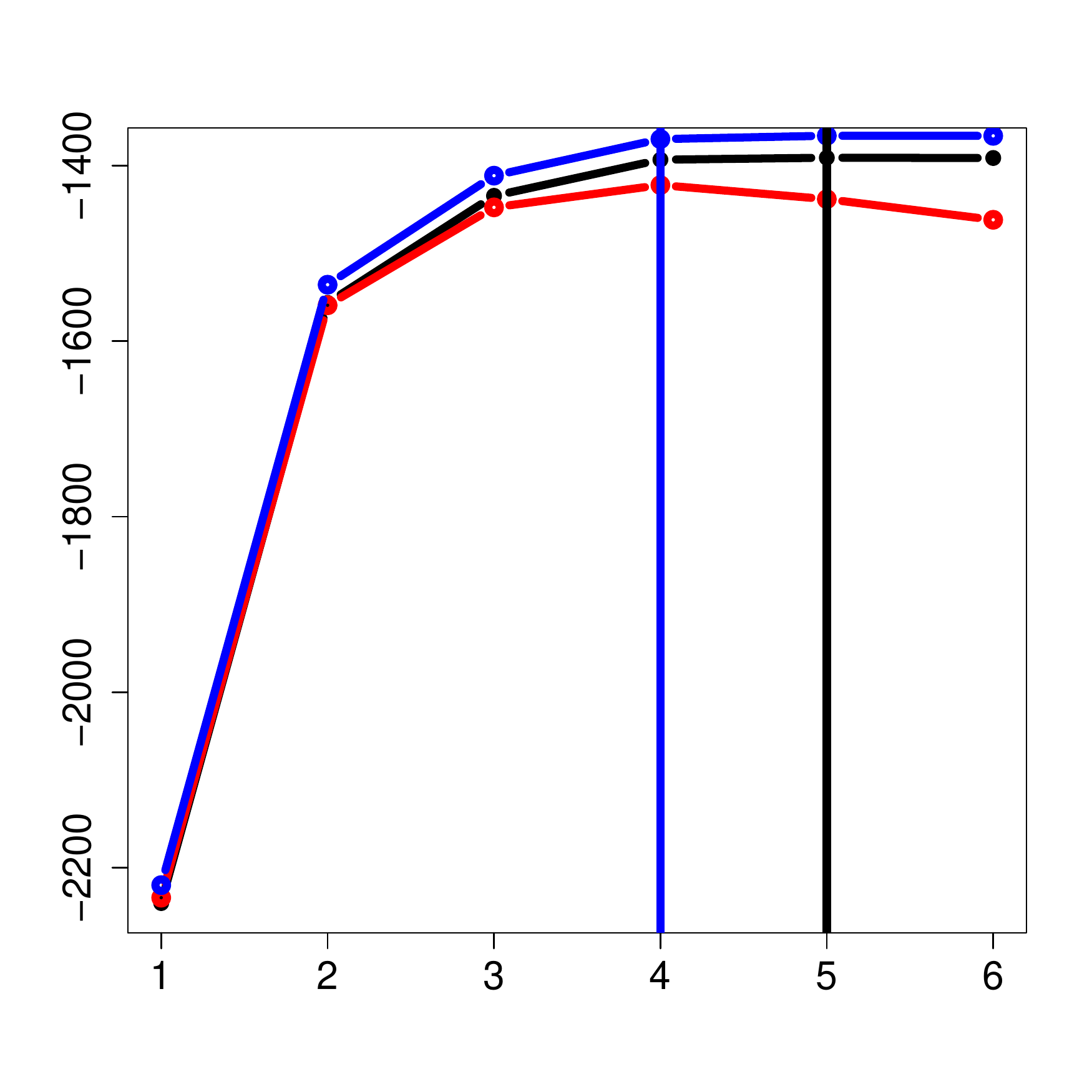} \end{tabular} &
   \begin{tabular}{c} \includegraphics[trim=10 10 10 10, width=.3\textwidth, height=.15\textheight]{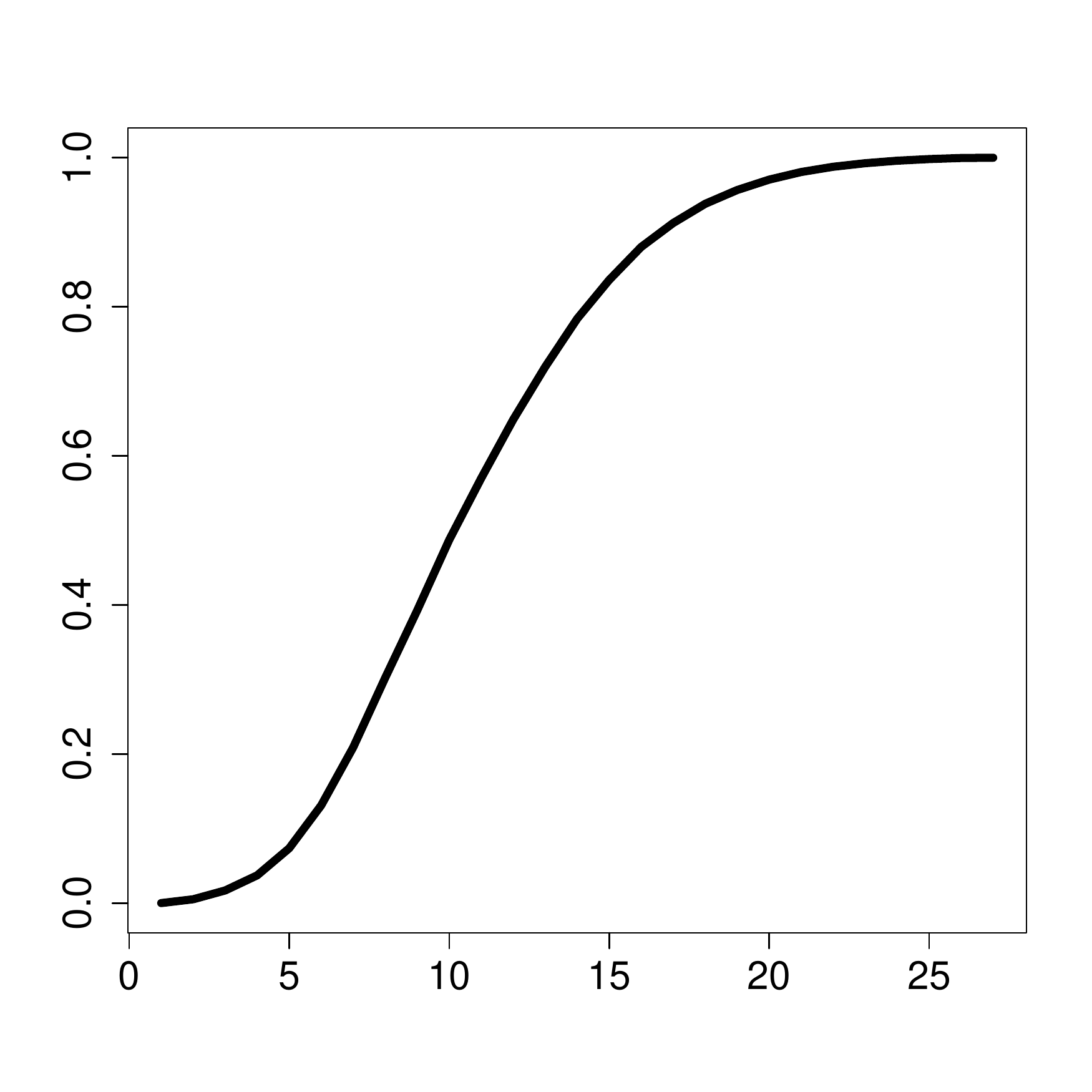} \end{tabular} & 
   \begin{tabular}{c} \includegraphics[trim=10 10 10 10, width=.3\textwidth, height=.15\textheight]{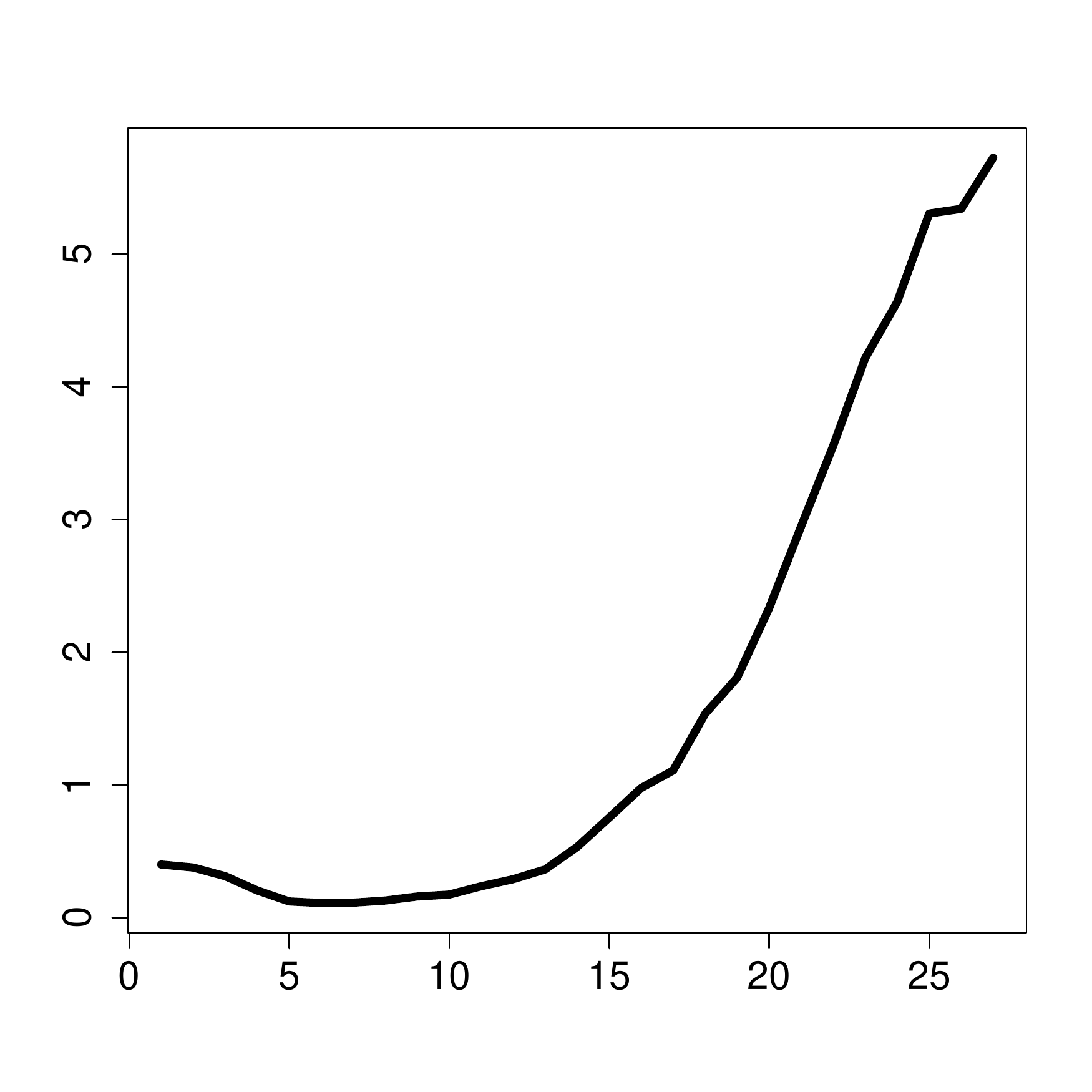} \end{tabular}  \\
   $\beta_1$ & $\beta_2$ & $\beta_3$ \\
   \begin{tabular}{c} \includegraphics[trim=10 10 10 10, width=.3\textwidth, height=.15\textheight]{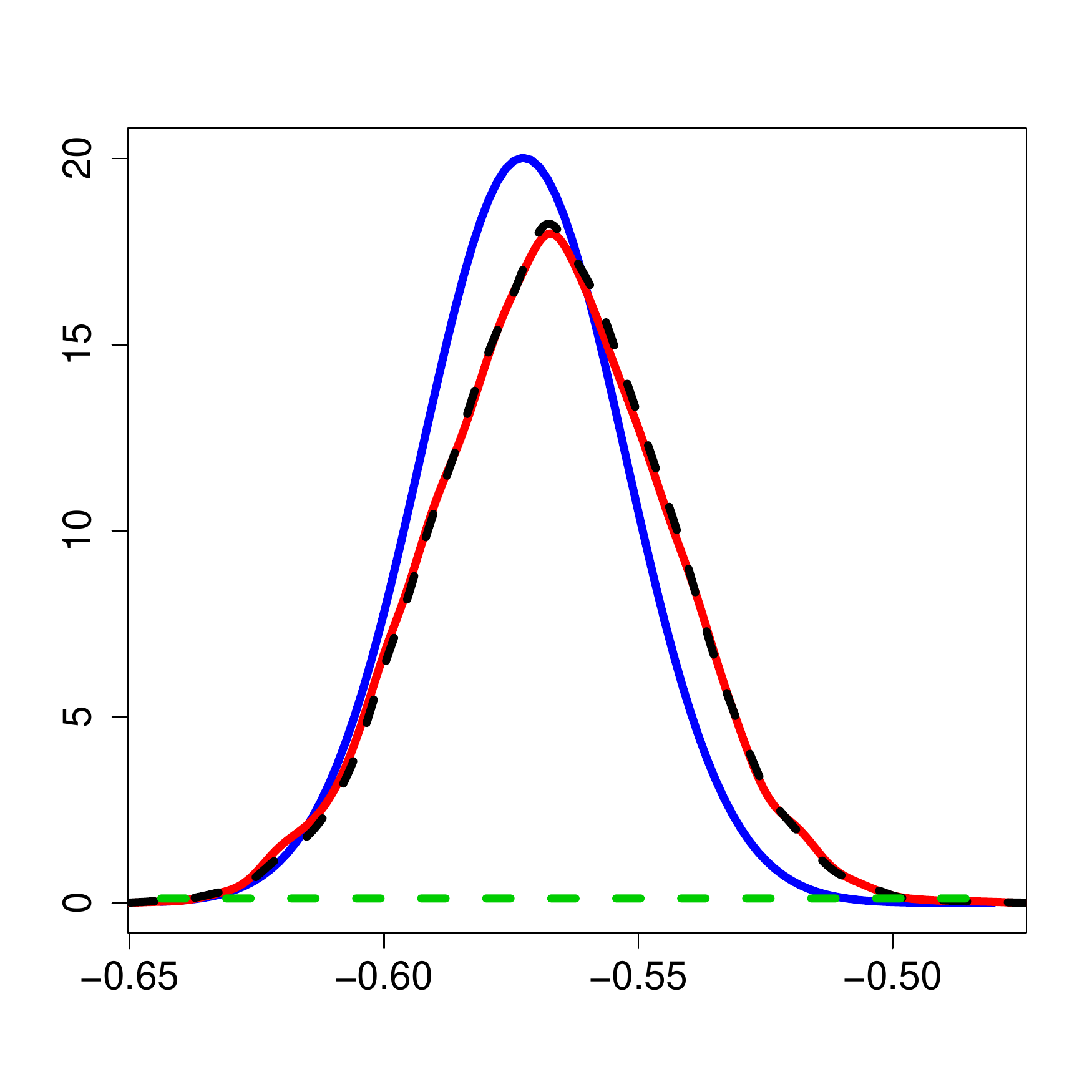} \end{tabular} &
   \begin{tabular}{c} \includegraphics[trim=10 10 10 10, width=.3\textwidth, height=.15\textheight]{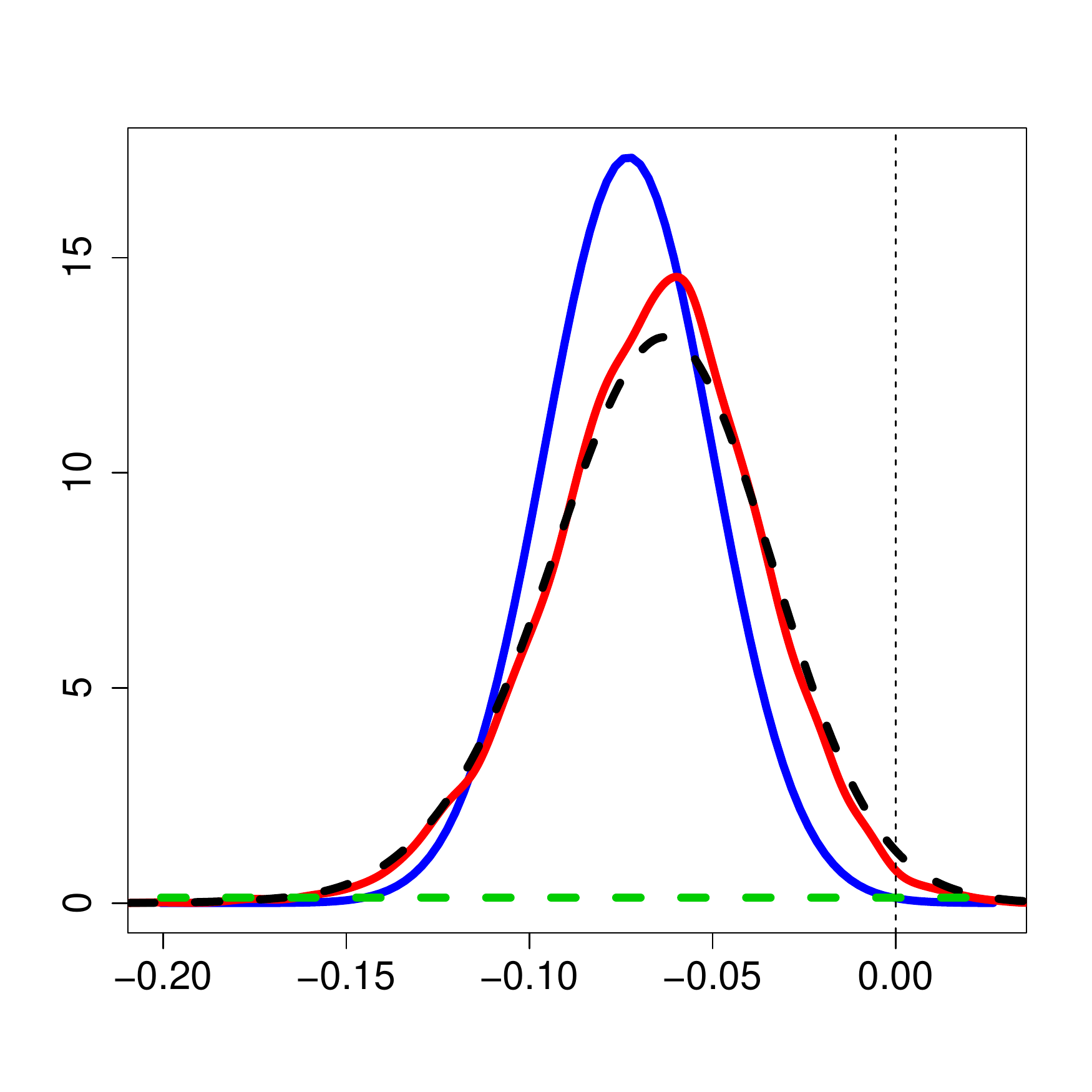} \end{tabular} &
   \begin{tabular}{c} \includegraphics[trim=10 10 10 10, width=.3\textwidth, height=.15\textheight]{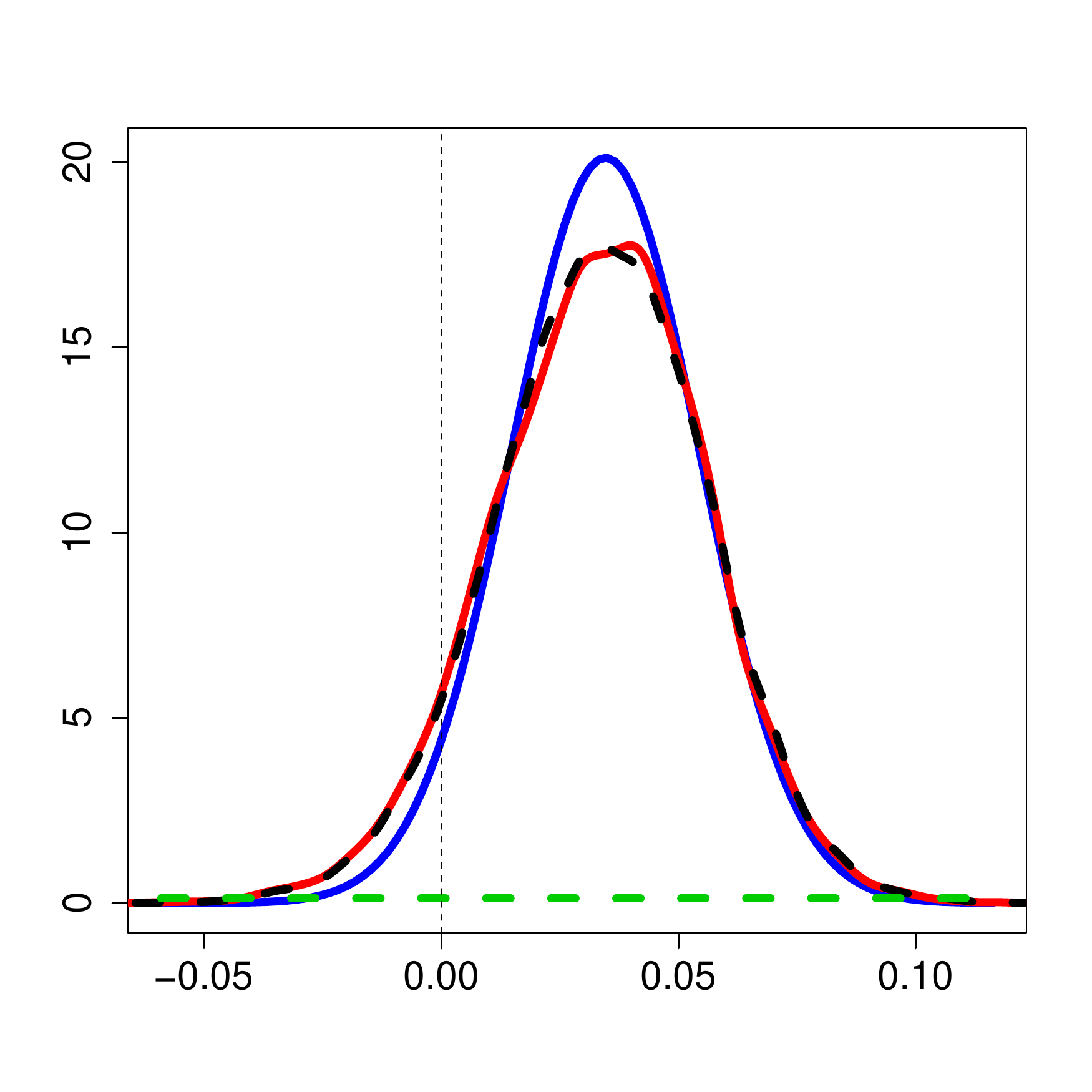} \end{tabular} 
  \end{tabular}
 \end{center}
 \caption{
 \emph{Results for the tree network.}
 Top left: selection of the number of clusters $K$ (blue: $J$ defined in equation \eqref{eq:J}, red: $ICL$, black: $\log p(Y)$).
 Top center: path of the bridge sampling.
 Top right: evolution of mutual information between the $Z_i$ along the sampling path.
 Bottom: posterior distribution of the regression coefficients $\beta_j$ for the taxonomic (left), the geographic (center) and the genetic (right) distance (dashed green: prior, blue: approximate posterior $\pt_Y(\beta)$, red: posterior  $p(\beta \mid Y, K)$ for the selected $K$, dashed black: posterior $p(\beta \mid Y)$, averaged over all $K$).
 \label{fig:tree}}
\end{figure}

\begin{table}
  \begin{tabular}{c|rrr}
    & 1--2 & 1--3 & 2--3 \\ \hline
    {$\papprox$} & -.037 & -.010 & .235 \\
    {\textsf{SMC from approx}} & -.139 & -.017 & .325
  \end{tabular}
  \label{tab:tree}
  \caption{Posterior correlation between the ${\beta}_j$ ('1-2': correlation between $\beta_1$ and $\beta_2$). } 
\end{table}

\begin{figure}[ht]
 \begin{center}
  \begin{tabular}{ccc}
   \includegraphics[trim=60 60 60 60, width=.3\textwidth]{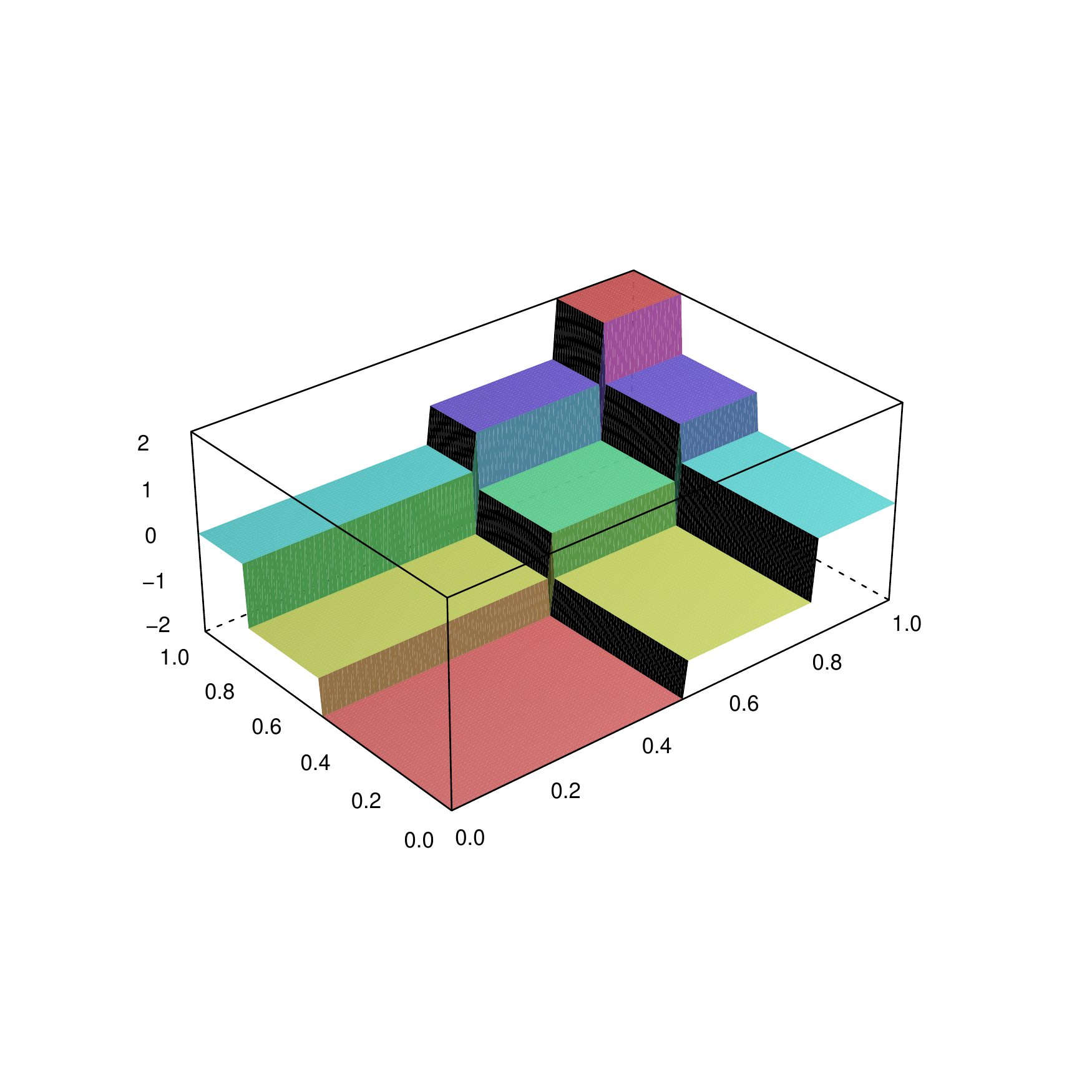} &
   \includegraphics[trim=60 60 60 60, width=.3\textwidth]{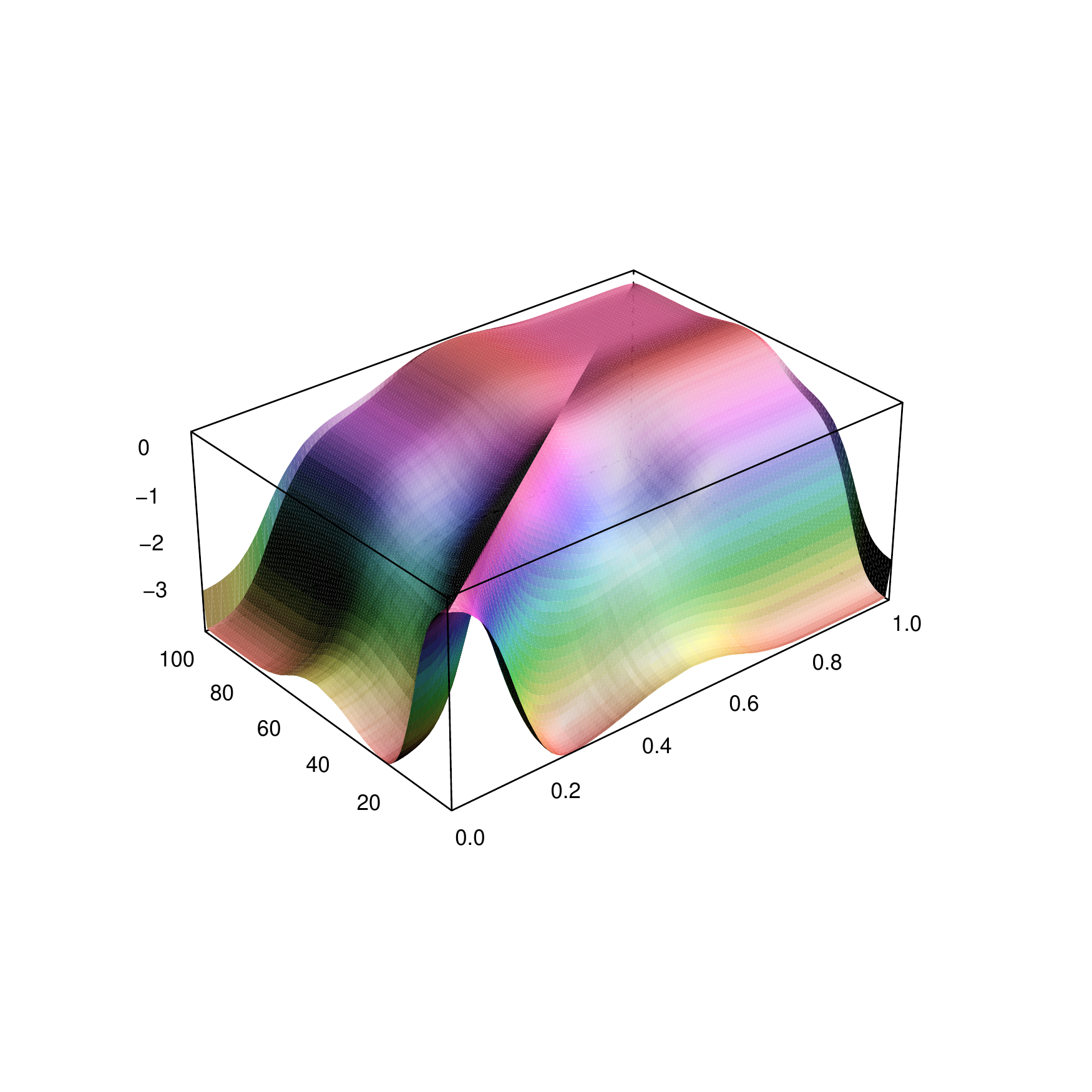} &
   \includegraphics[trim=60 60 60 60, width=.3\textwidth]{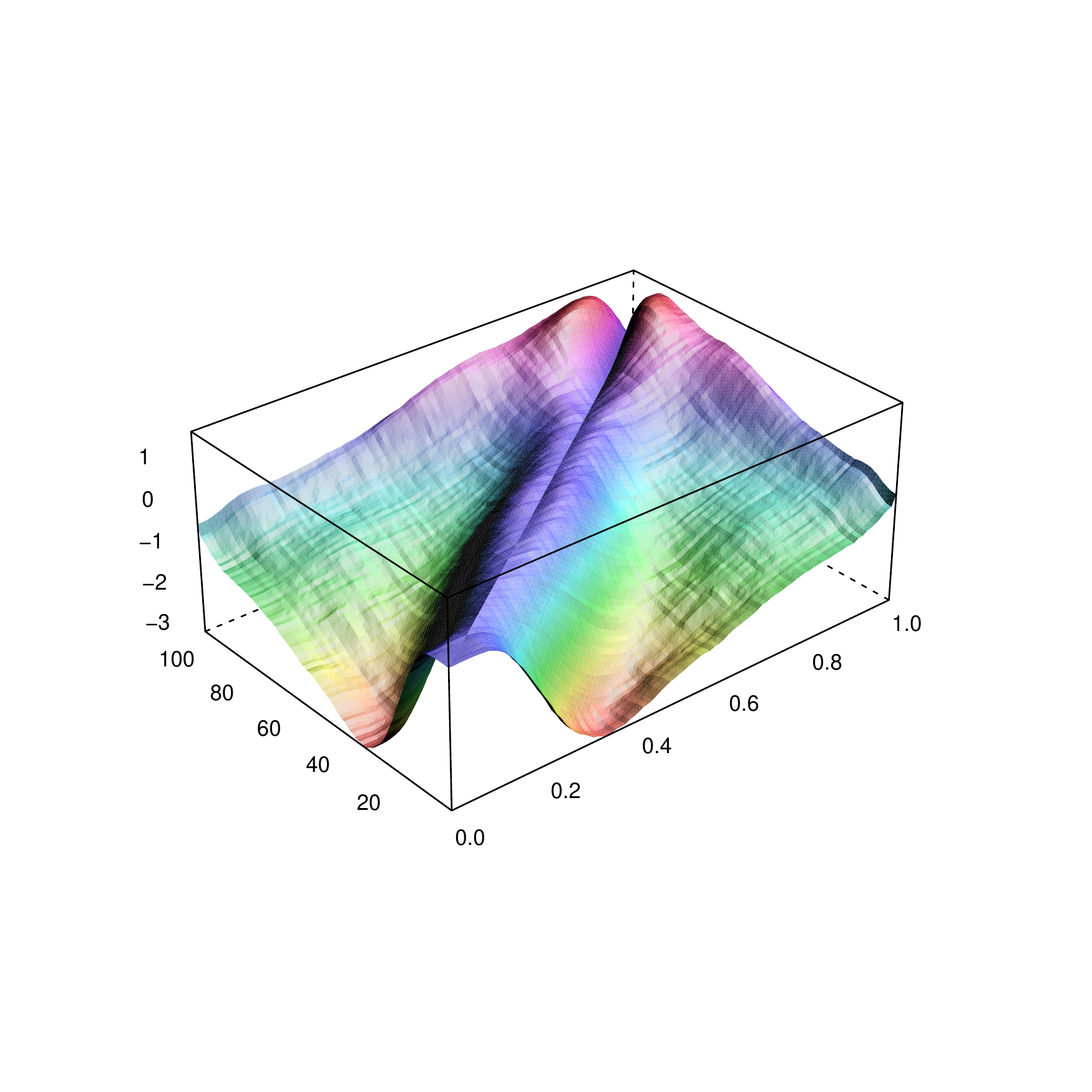} \\
   &
   \includegraphics[width=.3\textwidth]{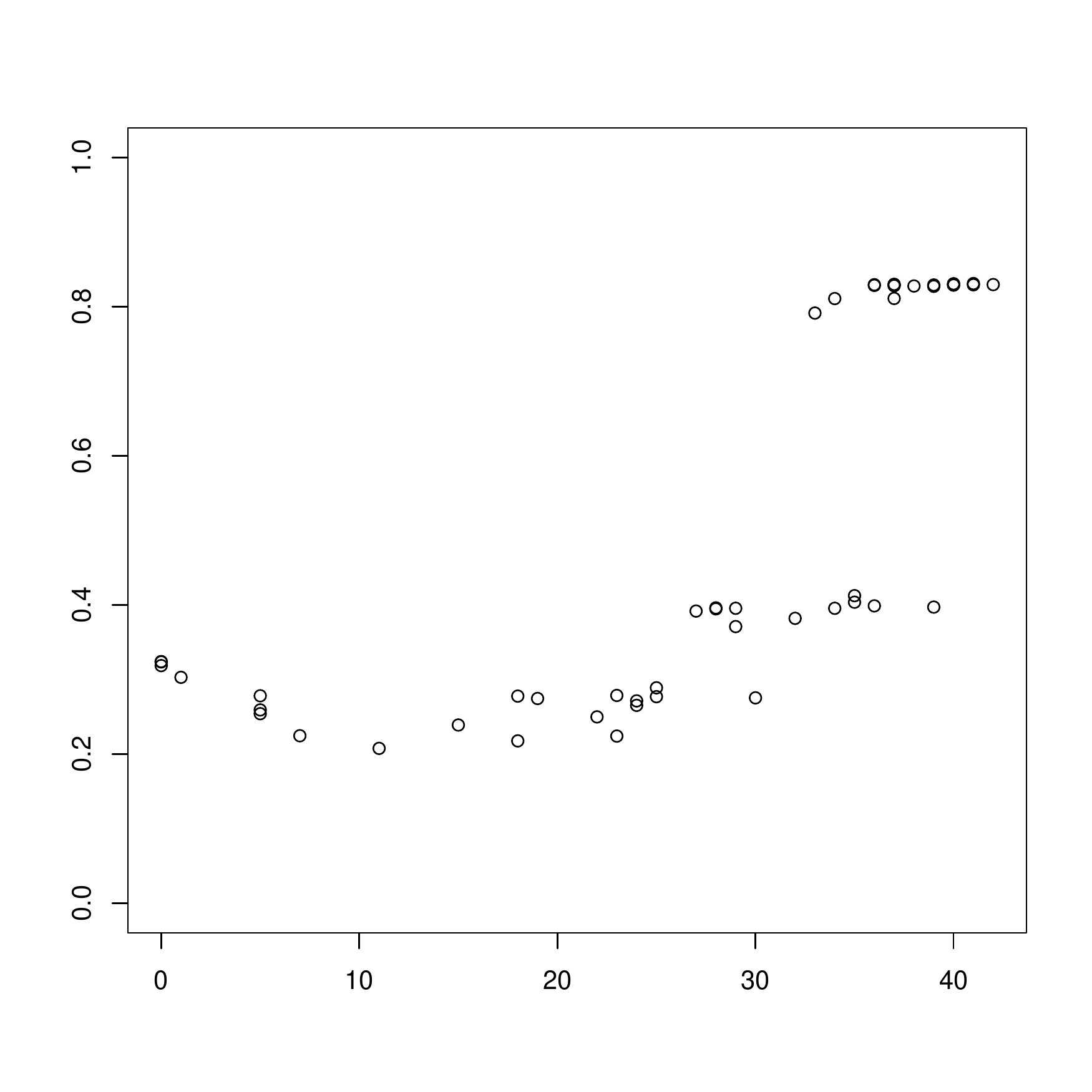} &
   \includegraphics[width=.3\textwidth]{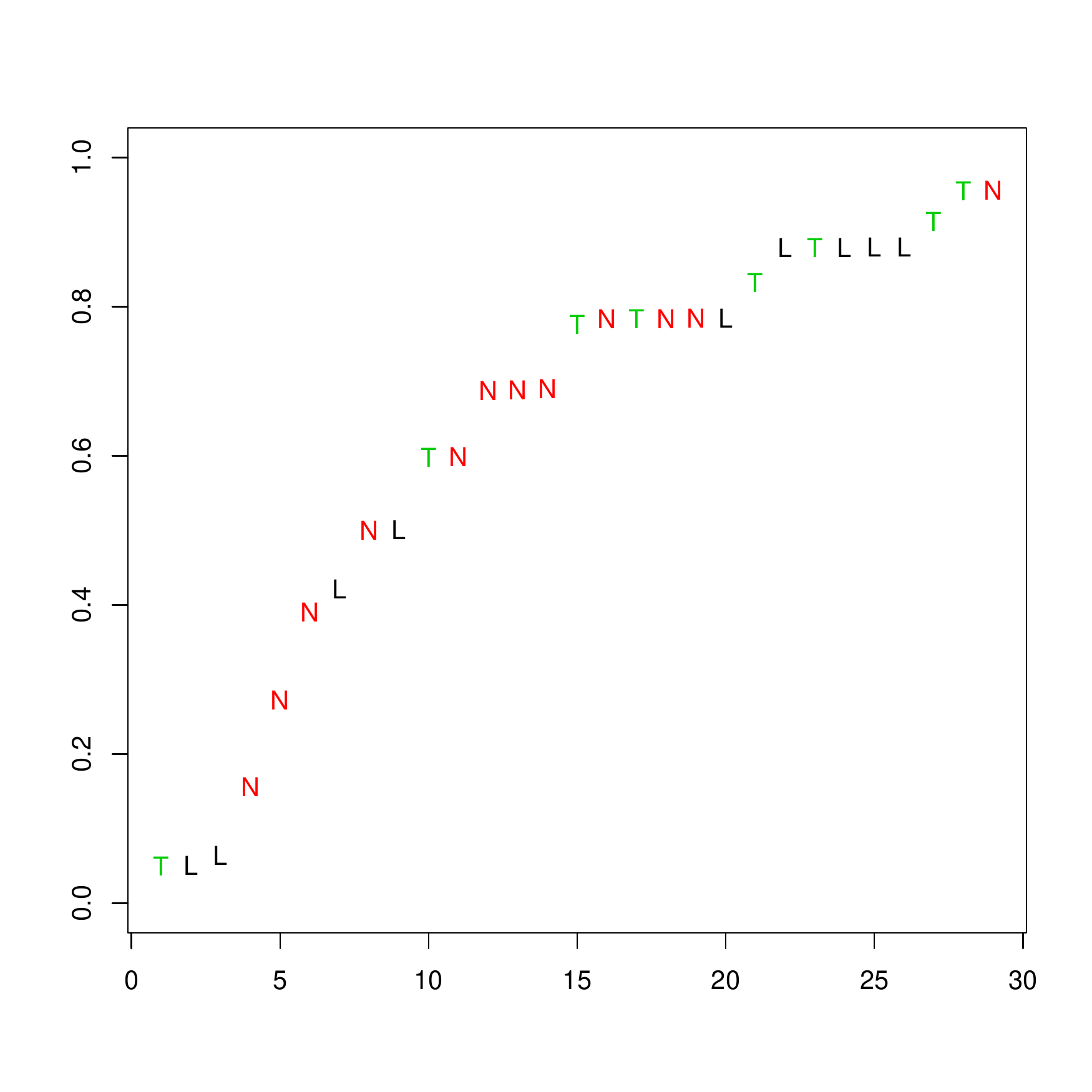} \\
  \end{tabular}
 \end{center}
 \caption{
 Top left: toy example of the functionfrom  $\phi$ of and \SBMreg model with $K=3$ groups.
 Top center: residual graphon for the tree network. 
 Top right: residual graphon for the onager network.
 Bottom center: $x$-axis = number of neighbours of node $i$ in the tree network, $y$-axis = posterior mean of $U_i$. 
 Top right: sorted posterior mean of the $U_i$, character = onager status (L, N or T).
 \label{fig:graphon}
 }
\end{figure}

The estimated residual structure $\widehat{\phi}$ for the tree network is displayed in Figure \ref{fig:graphon} (top center). The function is fairly flat but there exist a substantial fraction of species (with low coordinate $u$ or $v$), for which the observed number of interactions is lower than expected according to the distances. Interestingly, Figure \ref{fig:graphon} (bottom center) shows that the species with lowest coordinates are not the three species that are isolated in the network (Figure \ref{fig:net}, left).

\subsubsection*{Equid social networks}

The second example arise from \cite{RSF15}, who carried parallel analyses of the social networks of two equid species: the Grevy zebra and the onager. For each of the species, the interactions between all pairs of individuals were recorded during several days (44 for the zebras and 82 for the onagers). Within each species, we considered the total number $Y_{ij}$ of interactions between $i$ and $j$ during the time period. The age and sex of each animal were also recorded, resulting in four status for the zebras ($s$: stallion, $b$: bachelor, $n$: non-lactating, $l$: lactating) and three for the onagers ($T$: territorial male, $N$: non-lactating, $L$: lactating). The complete dataset is available on Dryad (\url{https://datadryad.org/resource/doi:10.5061/dryad.q660q}). \\
One question of interest here is to understand if the status of the individuals contributes to shape the interaction network. To better decompose this effect we decomposed the status as a combination of sex (($n$, $l$) vs ($s$, $b$) for the zebras and ($N$, $L$) vs ($T$) for the onagers) and age (($n$, $b$) vs ($s$, $l$) for the zebras and ($N$) vs ($T$, $L$) for the onagers). Then, for each pair we defined three binary variables indicating whether the two individuals share the same sex ($x^1$), the same age ($x^2$) or the same status ($x^3$). The most complete models are ($x^1, x^2, x^3$) for the zebras and ($x^1, x^2$) for the onagers, respectively. The posterior distribution of the corresponding regression coefficients are consistent with this conclusion (not shown).

Model comparison clearly shows that the sex ($x^1$) is the only significant effect for the zebra network (model posterior probability $=98.2\%$), whereas the combination of the sex and the age ($x^1, x^2$) contributes to structure the onager network (model posterior probability $\simeq 100\%$). This observation is consistent with the conclusion of \cite{RSF15}.

As an illustration, we further investigated the residual structure of the onager network. The estimated function $\widehat{\phi}$ (Figure \ref{fig:graphon}, top right) displays a complex pattern. A general trend from low to high coordinate suggests an heterogeneous propensity to social interaction between individuals. Figure \ref{fig:graphon} (bottom right) shows that no relation can be established between the latent coordinate $U_i$ of each individual and its sex or age, which confirms the residual nature of this structure with respect to the animal's status.

As a general comment about the proposed sampling procedure, we observed that few tens of steps are enough to shift from the variational-based approximate distribution to the exact posterior.

\section{Discussion}
In this work, we propose a proxy for the posterior  distribution of the Poisson SBM-reg model. This approximation 
is derived from the  combination of the {Variational EM} estimate and an approximation of the posterior variances using a  Laplace approximation.   Although this proxy for the posterior is a satisfactory approximation of the true posterior distribution  for some parameters of the model, 
it does not handle the posterior dependencies induced by the latent variables. We prove that this proxy can be used to supply an efficient sampler of the true posterior distribution. 

For the two motivating examples, the proposed procedure enables us to assess the effect of some of the covariates on the organisation of the interaction network, as well as the exitence of a residual structure that is not associated with these covariates.
 
Although the paper focuses on the Poisson SBM-reg model, our method is quite general. Indeed,  for any statistical method, provided one is able to build a proxy for the posterior distribution, SMC allows to use this proxy to sample efficiently  from the true posterior distribution. Variational Bayes or Expectation Propagation method are typical methods supplying an approximation of the posterior distribution. 
Applying our method starting from these approximations not only generates the expected sample but is also an indicator of how good the proxy is. A small number of iterations in the SMC   indicates that the proxy is an accurate  approximation of the posterior distribution.  

Moreover, deriving a Variational Bayes estimate is not always an easy task.   The proposed method to  build a proxy from the {Variational EM} can be straightforwardly  extended to any latent variable model  of the one-parameter exponential family. 

\paragraph{Acknowledgements.}
The authors thanks M. Mariadassou for helpful discussions.
This work has been supported by the French ANR projects EcoNet (ANR-18-CE02-0010) and ABSint (ANR-18-CE40-0034).

\bibliographystyle{chicago}
\bibliography{biblio}

\appendix
\section{Appendix}
\subsection{Details on the proxy of the posterior \label{app:proxy}}
 
\paragraph{Expression for $\partial^2_{\btheta^2} J(Y; \widetilde{\btheta}, \qt)$ }
Because $J(Y;  \btheta , \qt)  = \Espt \log p_\btheta(Y, Z)  +  \mathcal{H}(\qt)$ has been detailed in equation \eqref{eq: calcJ}, the derivation of the elements of $ \partial^2_{\btheta^2} J(Y; \btheta, \qt) =  \Espt \left(\partial^2_{\theta^2} \log p_{\btheta}(Y, Z)\right)  $ raises no specific difficulty. \\
Denoting $\widetilde{N}_k = \sum_i \tautik$ and $\etaijkl = \alphakl + \xij^\intercal \beta$, it suffices to remind that $\nuK = 1 - \sum_{k=1}^{K-1} \nuk$ and that $\Espt(\Zik \Zjl) = \tautik \tautjl$ (because $\qt$ is factorisable) to get
\begin{align}  \label{eq:d2.logpYZ}
  \text{for } 1 \leq k < K  \qquad \Espt \left(\partial^2_{\nuk,\nuk} \log p_\theta(Y, Z)\right) & = - \widetilde{N}_k / \nuk^2 - \widetilde{N}_K / \nuK^2, \nonumber \\
  \text{for } 1 \leq k \neq \ell  < K \qquad \Espt \left(\partial^2_{\nuk, \nul} \log p_\theta(Y, Z)\right) & = - \widetilde{N}_K / \nuK^2, \nonumber \\
  \text{for } 1 \leq k, \ell  < K \qquad \Espt \left(\partial^2_{\alphakl,\alphakl} \log p_\theta(Y, Z)\right) & = - \sum_{i < j} \tautik \tautjl e^{\etaijkl}, \\
  \Espt\left(\partial^2_{\beta,\beta} \log p_\theta(Y, Z)\right) & =  - \sum_{i < j} \xij \left(\sum_{k, \ell} \tautik \tautjl e^{\etaijkl} \right) \xij^\trans, \nonumber \\
  \Espt\left(\partial^2_{\beta, \alphakl} \log p_\theta(Y, Z)\right) & =  - \sum_{i < j}  \tautik \tautjl e^{\etaijkl}\xij, \nonumber 
\end{align}
all other second derivatives being null. As a consequence, the Hessian matrix $\partial^2_{\theta^2} J(Y; \theta, \qt)$ is made of two diagonal blocks corresponding to $\nu$ and $\gamma = (\alpha, \beta)$, respectively.

\subsection{Details on the SMC algorithm}\label{app:SMC}

We give here the details of the SMC algorithm developed by  \cite{DelMoral2006}  and  implemented in our procedure. 
  
\noindent Let us introduce the following notations: 
 \begin{eqnarray}\label{eq:gammah_Zh}
 \gamma_h(\bZ,\btheta) & = & \left[\papprox(\bZ, \btheta)\right]^ {1-\rhoh} \left[ \pi(\btheta) p_{\btheta}(\bZ) p_{\bZ,\btheta}(\bY) \right] ^{\rho_h},
\end{eqnarray}
and 
$ \Gamma_h = \int \gamma_h(\bZ,\theta) \d \theta \d \bZ \nonumber $, 
so that $\ph(\bZ,\btheta) = \gamma_h(\bZ,\btheta) / \Gamma_h$ is a probability density.
The main idea of \cite{DelMoral2006} is to plunge the problem of sampling a sequence of distributions defined on a single set $\Zcal \times \Theta$ into the standard SMC filtering framework. To that purpose, the sequence $(\ph)_{h = 0\dots H}$ is replaced by a sequence of extended distributions:

 \begin{equation}\label{eq:pibarh}
\pbar_h(\bthetazh) = \frac{\gammabar_h(\bthetazh)}{\Gamma_h}
\end{equation}
 with
 \begin{equation}\label{eq:pibarh2}
\gammabar_h( \bthetazh ) = \gamma_h(\bthetah) \prod_{k = 1}^{h} L_{k}\left(\bthetakmoins | \bthetak\right) 
\end{equation}
where $(\bthetazh) = (\bthetaz,\dots, \bthetah) \in\Zcal \times \Theta \times \dots\times \Zcal \times \Theta  = (\Zcal \times \Theta)^{h+1}$ and $(L_k)_{k = 0,\dots H-1}$ is a sequence of backward kernels satisfying: 
\begin{equation}\label{eq:L}
\int L_k\left(\bthetakmoins | \bthetak\right)d (\bthetakmoins) = 1, \quad \forall k = 0\dots H-1. 
\end{equation}
Due to identity \eqref{eq:L}, the marginal version of $\pbar_h$ (i.e. when integrating out $\bthetaz$ , $\dots$, $\bthetahmoins$) is the distribution of interest $\ph$. Once defined the sequence $(\pbar_h)_{h = 0\dots H}$, one may use the original SMC algorithm designed by \cite{Doucet2001} for filtering. At iteration $h$, the SMC sampler involves three steps: 
\begin{itemize}
\item \emph{Moving the particles} from $(\bthetahmoins)$ to $(\bthetah)$ using a transition kernel $ K_h(\bthetah | \bthetahmoins)$. As a consequence, let $ \eta_{h-1}(\bthetazhmoins)$ denote the sampling kernel for $ (\bthetazhmoins)$ until iteration $h-1$, $\eta_h$'s expression is:
\begin{equation}\label{eq:eta}
 \eta_h(\bthetazh) = \eta_{h-1}(\bthetahmoins) K_h(\bthetah | \bthetahmoins)
 \end{equation}
 \item \emph{Reweighing the particles} in order to correct the discrepancy between the sampling distribution $\eta_h$ and the distribution of interest at iteration $h$, $\pbar_h$. 
\item \emph{Selecting the particles} in order to reduce the variability of the importance sampling weights and avoid degeneracy. In practice the particles will be resampled when the  $ESS$  decreases below a pre-specified rate. 
\end{itemize}

\paragraph{About the importance weights.} 
 At iteration $h$, the importance sampling weights for $(\bthetazhm)_{m = 1\dots M}$ are : $\forall m = 1\dots M$,
\begin{equation}\label{eq:w1}
w_h^m = \wh(\bthetazhm) = \frac{\gammabar_h(\bthetazhm)}{\eta_h(\bthetazhm)}
\end{equation} 
in their un-normalized version. $(W^m_h)_{m = 1\dots M }$ denotes the normalized weights, i.e. 
\begin{equation*}
\Wh^m = \frac{\whm}{\ \sum_{m' = 1}^M w^{m'}_h}, \quad \forall m = 1\dots M
\end{equation*}
Equations (\ref{eq:pibarh}-\ref{eq:pibarh2}-\ref{eq:eta}-\ref{eq:w1}) imply a recurrence formula for the weight of any particle $(\bthetazh)$: 
\begin{equation}\label{eq:w}
\wh(\bthetazh) = w_{h-1}(\bthetazhmoins) \wtilde_{h-1:h}(\bthetahmoins,\bthetah)
\end{equation}
where the incremental weight $ \wtilde_{h-1:h}(\bthetahmoins,\bthetah)$ is equal to: 
 \begin{equation}\label{eq:wtilde}
 \wtilde_{h-1:h}(\bthetahmoins,\bthetah) = \frac{L_h(\bthetahmoins | \bthetah)}{ K_h(\bthetah | \bthetahmoins)}\frac{\gamma_h(\bthetah)}{\gamma_{h-1}(\bthetahmoins)}
\end{equation}

\paragraph{About the transition kernels $K_h$  and $L_h$.} As, at this step, the target distribution is $\ph$, it seems natural to choose $K_h(\bthetah | \bthetahmoins)$ as a Monte Carlo Markov Chain (MCMC) kernel with $\ph(\bZ,\btheta)$ 
as stationary distribution. 
Following \cite{DelMoral2006}, we choose the backward kernel: 
\begin{equation}\label{eq:L2}
L_h(\bthetahmoins | \bthetah) = \frac{ K_h(\bthetah | \bthetahmoins) p_{h}(\bthetahmoins) }{\ph(\bthetah)} 
\end{equation}
which satisfies Property \eqref{eq:L} and enables us to rewrite the weight increment $ \wtilde_{h-1:h}(\bthetahmoins,\bthetah)$ appearing in \eqref{eq:w} and defined in \eqref{eq:wtilde} as
 \begin{equation}\label{eq:wtilde2}
 \wtilde_{h-1:h}(\bthetahmoins,\bthetah) = \frac{\gamma_{h}(\bthetahmoins) }{\gamma_{h-1}(\bthetahmoins)} = \left[r(\bthetahmoins)\right]^{\rho_h-\rho_{h-1}}
\end{equation}
where
\begin{equation}\label{eq:alpha}
r(\bZ,\btheta) = \frac{p_{\btheta}(\bY,\bZ) \pi(\btheta)}{\qapprox(\bZ,\btheta| \bY)}. 
\end{equation}
In what follows, we denote 
\begin{equation}\label{eq:def r}
r _h = r(\bZ_h^m,\btheta_h^m)   \frac{p_{\btheta_h^m}(\bY,\bZ_h^m) \pi(\btheta_h^m)}{\qapprox(\bthetahm \mid  \bY)}. 
\end{equation}

\begin{Remark}
Using this particular backward kernel \eqref{eq:L2} has two major consequences. First it is not required having an explicit expression for the transition kernel $K_h(\btheta_h | \btheta_{h-1})$, which is quite welcome for MCMC kernels. Secondly, examining equations \eqref{eq:w} and \eqref{eq:wtilde2}, one may notice that the weight for a particle $\bthetazh$ does not depend on $\bthetah$ but only on $\bthetahmoins$. As a consequence, the weights of the particles $\bthetazh$ can be computed before they are simulated and for any new $\ph$
\end{Remark}

\paragraph{Adaptive design of $(\rho_h)_{h = 0\dots H}$.} 
As a consequence of this last remark, we are able to design an adaptive strategy for $(\rhoh)_{h = 0, \dots H}$ \cite[as in][]{Schafer2013,Jasra2011}. Indeed, being able to compute the weights of the up-coming particles for any new $\rho_h$, we can increase $\rho_h$ until the quality of the sample (measured through an indicator computed from the weights) decreases for the next distribution. In practice, 
following \cite{Zhou2016}, we use the conditional Effective Sampling Size ($cESS$) to measure the quality of $p_ {h-1}$ as an importance sampler when estimating an expectation against $\ph$. It is defined as: 
 \begin{eqnarray*}\label{eq:cESS}
 cESS & = & \left[\ \sum_{m = 1}^M M W_{h-1}^m \left(\frac{ \wtilde_{h-1:h}^m}{\ \sum_{m = 1}^M MW_{h-1}^m \wtilde_{h-1:h}^m} \right)^2 \right]^{-1} \\
 & = & \frac{M \left(\ \sum_{m = 1}^M W_{h-1}^m \wtilde_{h-1:h}^m\right)^2}{\ \sum_{m = 1}^M W_{h-1}^m ( \wtilde_{h-1:h}^m)^2}, 
\end{eqnarray*}
becoming 
\begin{eqnarray}\label{eq:cESS2}
&&cESS\left(\rho_h;\rho_{h-1}, (W_{h-1}^m,r_{h-1}^m )_{m  \leq  M}\right) =cESS_{h-1}(\rho_h)\nonumber\\
&&
= \frac{M \left(\ \sum_{m = 1}^M W_{h-1}^m (r_{h-1}^m)^{\rho_h-\rho_{h-1}}\right)^2}{\ \sum_{m = 1}^M W_{h-1}^m (r_{h-1}^m)^{2(\rho_h -\rho_{h-1})}}. 
\end{eqnarray}
where $r_{h-1}^m$ as been defined in Equation \eqref{eq:def r}. 
If $\rho_h = \rho_{h-1}$ , $cESS$ is maximal (equal to $M$, the number of particles). As $\rho_h$ increases, the discrepancy between $p_ {h-1}$ and $\ph$ increases and so the quality of $p_ {h-1}$ as an importance sampling distribution when estimating an expectation against $\ph$ decreases and so does $cESS$. As a consequence, our strategy to find the next $\rho_h$ is to set: 
$$ 
 \rho_ h = 1 \wedge \sup_{\rho}\left\{\rho > \rho_{h-1}, cESS_{h-1}(\rho) \geq \tau_1 M \right\}
 $$

\paragraph{Selection of the particles.} In order to prevent a degeneration of the particle approximation, we use a standard resampling of the particles. In other words, if the variance of weights $(W_{h}^m)_{m = 1\dots M}$ is too high (or in other words, if the $ESS$ is too small), we resample the particles using a multinomial distribution, thus discarding the particles with low weights and duplicating the particles with high weights. 


\paragraph{Sampling algorithm.} 
Finally, each iteration $h$ of the sequential sampler algorithm  consists in performing  the   steps "find $\rho_h$",  "resample",  "move the particles" and  "compute the new weights",  resulting into the algorithm described at page \pageref{page:algo}.

\paragraph{Estimation of the marginal likelihood }

\noindent Let us recall that $\Gamma_ h = \int  \gamma_h(\bZ,\btheta) \d\bZ \d\btheta$  has been defined at equation \eqref{eq:gammah_Zh}. Following \cite{DelMoral2006} and using the notations introduced  before, the ratio of the quantities  ${\Gamma_{h}} / {\Gamma_{h-1}}$ is estimated by: 
\begin{equation*}
\widehat{\frac{\Gamma_{h}}{\Gamma_{h-1}}} = \sum_{m = 1}^M \Whm \wtilde_{h-1:h}^m. 
\end{equation*}
and 
\begin{equation}\label{eq:pZhat}
\widehat{p}_{\bY}  = \widehat{\frac{\Gamma_{H}}{\Gamma_{0}}} = \prod_{h = 1}^H \widehat{\frac{\Gamma_{h}}{\Gamma_{h-1}}} = \prod_{h = 1}^H \sum_{m = 1}^M \Whm \wtilde_{h-1:h}^m 
\end{equation}
is naturally an  estimator of $\Gamma_H/Z_0$. However, 
\begin{eqnarray*}
\Gamma_H &=&   \int \gamma_H(\bZ,\theta)  \d \bZ\d \theta =  \int  \pi(\btheta) p_{\btheta}(\bZ) p_{\bZ,\btheta}(\bY)  \d \bZ  \d \theta  = p(\bY)
\end{eqnarray*}
and $\Gamma_0 = \int \gamma_0(\bZ,\theta)  \d \bZ \d \theta =\int \papprox(\bZ, \btheta)  \d \bZ \d \theta = 1$. As a consequence,  $\widehat{p}_{\bY}$ is an estimator of $p(\bY)$. 
\\

\noindent Note that another estimate is given by the path sampling identity. Indeed, under non-restrictive regularity assumptions, the following equality holds: 
\begin{equation}\label{eq:Int}
\log p(Y)  = \int_{0}^{1}  \Esp_{p_{\rho}}\left[\frac{d \log \gamma_{\rho}(\cdot)}{d \rho} \right]d\rho
\end{equation} 
where $\gamma_\rho(\bZ,\btheta) =\left[\papprox(\bZ, \btheta)\right]^ {1-\rho} \left[ \pi(\btheta) p_{\btheta}(\bZ) p_{\bZ,\btheta}(\bY) \right] ^{\rho}$,  and $p_\rho()$ is the associated probability density distribution and, in our geometric path sampling:
$$\frac{d \log \gamma_{\rho}(\bZ,\btheta)}{d \rho} = \log \frac{ \pi(\btheta) p_{\btheta}(\bZ) p_{\bZ,\btheta}(\bY)}{\qapprox(\btheta)} = \log r(\theta)$$
 An elementary trapezoidal scheme and Monte Carlo approximations of the expectations involved in \eqref{eq:Int} lead to the following approximation of the marginal likelihood:
\begin{equation}\label{eq:pZhat2}
\widehat{\widehat{\log p(\bY)}} = \sum_{h = 1}^H \frac{\rho_h - \rho_{h-1}}{2} (U^M_h + U^M_{h-1})
\end{equation}
where
 $U^M_h = \widehat{  \Esp}_{p_{\rho_h}}\left[ \log r(\theta)\right] = \sum_{m = 1}^M W_{h}^m \log r^m_{h}
 $. 

\begin{Remark}
Note that as suggested in \cite{Zhou2016}, we noticed on simulation studies that the two estimators behave similarly in our examples.  A precise comparison of the two estimators is out of the scope of this paper. 
\end{Remark}

\subsection{$\Phi$ functions used to illustrate the validity of our method}\label{sec append phi}

We test the validity of out method using the following eleven functions $\Phi$. 
\begin{equation*}
\begin{array}{lcllcl}
\Phi_1(\theta)&=& \sum_{r=1}^4\beta_r, \quad & \Phi_2(\theta) &=&|\pi_1 - \pi_2| ,  \\
\Phi_3(\theta) &=&\beta_1, \quad&\Phi_4(\theta)  &=&\beta_2,\\
\Phi_5(\theta) &=&\beta_3, \quad&\Phi_6 (\theta) &=&\beta_4 ,\\ 
\Phi_7(\theta) &=&  \alpha_{11}  + \alpha_{22} , \quad& \Phi_8(\theta) &=& \sum_{k,\ell} \alpha_{k\ell}, \\ 
\Phi_9(\theta) &=& \sum_{k,\ell} \alpha_{k\ell}+ \beta_1 , \quad& \Phi_{10 }(\theta)&=&  \sum_{k,\ell} \alpha_{k\ell}+ \beta_2 ,\\
\Phi_{11 }(\theta)&=&   \sum_{k,\ell} \alpha_{k\ell}+ \beta_3 , \quad&\Phi_{12 }(\theta)&=&    \sum_{k,\ell} \alpha_{k\ell}+ \beta_4,\\
\Phi_{13 }(\theta)&=&   \alpha_{11}  + \alpha_{22}  +  \sum_{r=1} \beta_r , \quad& \Phi_{14}(\theta) &=& \sum_{k,\ell} \alpha_{k\ell}  + \sum_{r=1}^4 \beta_r + |\pi_1 - \pi_2|  
 \end{array}. 
\end{equation*}

\end{document}